\documentclass[smallextended]{svjour3}      

\smartqed  

\usepackage[utf8]{inputenc} 
\usepackage{amsmath,amssymb}
\usepackage{latexsym}
\usepackage{graphicx}
\usepackage{multirow}
\usepackage{float}
\usepackage{subcaption}
\usepackage{float,color}

 \newcommand{\stdp}{\strut\displaystyle}

\begin{document}

\title{Entropy-based time-varying window width selection for nonlinear-type time-frequency analysis}

\titlerunning{TVOWW and AOWW}

\author{Yae-Lin~Sheu \and Liang-Yan~Hsu \and Pi-Tai~Chou \and Hau-Tieng~Wu}

\institute{Y.-L. Sheu \at the Department of Chemistry, National Taiwan University, Taipei, Taiwan and the Department of Mathematics, University of Toronto, M5S 2E4, Toronto, Canada 
	\and
L.-Y. Hsu \at the Department of Chemistry, Northwestern University, 60208, Evanston, USA
	\and
P.-T. Chou \at the Department of Chemistry, National Taiwan University, Taipei, Taiwan
	\and
H.-T. Wu \at the Department of Mathematics, University of Toronto, M5S 2E4, Toronto, Canada and Mathematics Division, National Center for Theoretical Sciences, Taipei, Taiwan 
\email{hauwu@math.toronto.edu}
}

\date{Received: date / Accepted: date}

\maketitle

\begin{abstract} 
We propose a time-varying optimal window width (TVOWW) scheme {and an adaptive optimal window width (AOWW)} selection scheme to optimize the performance of several nonlinear-type time-frequency analyses, including the reassignment method and its variations.
A window rendering the most concentrated distribution in the time-frequency representation (TFR) is regarded as the optimal window.
The TVOWW selection scheme is particularly useful for signals that comprise fast-varying instantaneous frequencies and small spectral gaps.
To demonstrate the efficacy of the method, in addition to analyzing synthetic signals, we study an atomic time-varying dipole moment driven by two-color mid-infrared laser fields in attosecond physics and near-threshold harmonics of a hydrogen atom in the strong laser field. 
\end{abstract}

\section{Introduction}

Scientists investigate nature by collecting diverse types of data.
They then infer the underlying rules by modeling and analyzing the recorded data. Time series is a commonly encountered data type. Its time-evolving nature paves the way for scientists to access the system's dynamics. Time-frequency (TF) analysis is a powerful time series analysis tool, which captures nonstationary oscillatory dynamics and serves as a portal to the underlying system.

During the past $70$ years, several TF analysis methods were developed \cite{Flandrin_Book:1999}, which can be classified into three types: linear, quadratic, and nonlinear.
Linear-type transforms, such as the short time Fourier transform (STFT) and the continuous wavelet transform (CWT), have been widely studied. 
They are subject to the limitation of the uncertainty principle associated with the CWT, or the STFT \cite{Flandrin_Book:1999,Grochenig:2001,Ricaud_Torresani:2014}.
Quadratic-type transforms, such as the Wigner-Ville distribution and Cohen class, could provide a more adaptive analysis of the input signal. However, they suffer from severe mode mixing artifacts \cite{Flandrin_Book:1999}.
There are several nonlinear-type transforms including: the reassignment method (RM) \cite{Chassande-Mottin_Auger_Flandrin:2003,Auger_Chassande-Mottin_Flandrin:2012} and its variations, the TF by convex optimization (Tycoon) \cite{Kowalski_Meynard_Wu:2015}, the Blaschke decomposition (BKD) \cite{Coifman_Steinerberger:2015,Coifman_Steinerberger_Wu:2016}, the empirical mode decomposition (EMD) \cite{Huang_Shen_Long_Wu_Shih_Zheng_Yen_Tung_Liu:1998}, the iterative filtering \cite{Cicone_Liu_Zhou:2014}, the sparsification approach \cite{Hou_Shi:2013a}, the approximation approach \cite{Chui_Mhaskar:2016}, the TF jigsaw puzzle (TFJP) for the Gabor transform (GT) \cite{Jaillet_Torresani:2007,Ricaud:2014}, the non-stationary GT (NSGT) \cite{Balazs_Dorfler_Jaillet_Holighaus_Velasco:2011}, the matching pursuit \cite{Mallat_Zhang:1993}, and several others. 
The variations of RM include: the synchrosqueezing transform (SST) \cite{Daubechies_Lu_Wu:2011,Wu:2011Thesis}, the synchrosqueezed wave packet transform \cite{Yang:2014}, the synchrosqueezed S-transform \cite{Huang_Zhang_Zhao_Sun:2015}, the second-order SST \cite{Oberlin_Meignen_Perrier:2015}, the concentration of frequency and time (ConceFT) \cite{Daubechies_Wang_Wu:2016}, and the de-shape SST \cite{Lin_Li_Wu:2016}. While the approaches vary from algorithm to algorithm, the common goal of nonlinear-type transforms is to obtain a ``sharpened'' TF representation (TFR) that could provide more accurate dynamical information underlying the recorded time series. We refer interested readers to \cite{Daubechies_Wang_Wu:2016} for a more extensive literature survey and their applications.

The nonlinear-type transforms can be classified into two categories. The first category consists of transforms that do not require choosing a window, like the BKD, the EMD, and the Tycoon. While the EMD has been widely applied, its application to data analysis needs more attention due to its lack of mathematical foundation. The BKD, on the other hand, is solidly supported by the complex analysis theory. However, there are still several mathematical challenges left unsolved, and the application of the BKD to data analysis is still in its infancy. The Tycoon is a synthesis-based approach to estimate the TFR with the sparsity constraint based on the convex optimization. The Tycoon theoretically has the potential to achieve a sharp TFR, but it is currently compute-intensive. 

The second category consists of transforms that depend on a chosen window, 
which can be classified into two subcategories: reassignment-type and non-reassignment-type. The reassignment-type subcategory includes the RM and its variations, and the non-reassignment-type subcategory includes the other algorithms. While different methods are subject to different limitations, they are all limited by the \textit{window selection} problem. The question is: what is the \textit{optimal} window when we analyze a given time series? In the ideal situation, the optimal window should be universal and always provides the optimal results under some constraints. However, it is widely believed that there probably is no optimal window due to the complicated nonlinearity hidden inside the natural signals. To resolve this issue, different methods provide different solutions. 
For example, in the reassignment-type transforms, we could theoretically prove that when the signal and window satisfy some regularity conditions, the algorithms are \textit{adaptive} to the signal, in the sense that the dependence on the window is negligible; see, for example, \cite[Theorem 3.3]{Daubechies_Lu_Wu:2011}. However, in practice, the situation might be more complicated. Therefore, the performance of the algorithm is not guaranteed. Thus, how to determine the optimal window for nonlinear time series is a crucial issue.

In this paper, we aim to alleviate this window selection issue for the reassignment-type transforms. 
We consider the R\'enyi entropy to determine the optimal window. By applying the optimal window width, the TFR sharpness can be enhanced while the reconstruction routine of the SST and its variations can be preserved.
We specifically consider a window that is \textit{optimal} for a chosen TF analysis, if the distribution of the associated TFR is highly concentrated. While there are several ways to measure the distribution concentration, we apply the R\'enyi entropy \cite{Coifman_Wickerhauser:1992,Baraniuk_Flandrin_Janssen_Michel:2001,Sedic_Djurovic_Jiang:2009}, which has been shown to efficiently estimate the signal information content and complexity in the TFR.

The article is organized as follows.
Section \ref{Section:gAHM} summarizes the background material, including the adaptive harmonic model (AHM) describing an oscillatory signal composed of multiple components, and several reassignment-type TF analysis tools that could be applied to analyze such signals.
Section \ref{Section:TVOW} describes a scheme to optimize the performance of the reassignment-type TF analyses by {window width selection techniques}. A comparison of the proposed scheme and some non-reassignment-type transforms is also provided.
Numerical results and an application to the attosecond physics are reported in Section \ref{Section:Results}.
A conclusion is drawn in Section \ref{Section:Conclusions}.

\section{Background}  \label{Section:gAHM}

In this section, we summarize the AHM to quantify oscillatory signals, and review several recently proposed TF analysis tools suitable for analyzing signals satisfying the AHM. While the review could be extended to other reassignment-type transforms, such as the RM, the de-shape SST and the ConceFT, we only review the SST\footnote{ The SST can be defined also on the CWT \cite{Daubechies_Lu_Wu:2011}, the S-transform \cite{Huang_Zhang_Zhao_Sun:2015}, as well as other linear-type TF transforms \cite{Yang:2014}. Here we focus only on the SST defined on the STFT.} and the 2nd-order SST in this section.

\subsection{Adaptive Harmonic Model}

The AHM aims to describe the time-varying oscillatory dynamics in a given signal.
Suppose that the signal $x(t)$ is composed of finite $K\geq 1$ oscillatory functions; that is, $x(t)=\sum_{k=1}^K f_k(t)$, where $f_k$ is the $k$-th oscillatory function and $k=1,\ldots,K$.
The $k$-th oscillatory function $f_k$ is composed of an amplitude modulation (AM) $a_k(t)$, which is positive, and a phase function $\phi_k(t)$, which is strictly monotonically increasing, so that $f_k(t)=a_k(t)\cos(2\pi\phi_k(t))$, for $k=1,\ldots,K$.
The $\phi'_k(t)$ is thus positive and is regarded as the instantaneous frequency (IF) of the $k$-th oscillatory function. In this study, we consider only real oscillatory signals, since most time series we acquire in the real world are real.

While such a AHM describes a signal composed of multiple oscillatory functions, it is too general to work with and we need some constraints. Fix $\epsilon\geq 0$. Let the positive constant $c$ be the supremum of the variation of the IF function; that is, $\|\phi_k''\|_{\infty}\leq c$ for $k=1,\ldots,K$. It is also assumed that the variation of the AM is controlled by the IF; that is, $|a_k'(t)|\leq \epsilon \phi'_k(t)$ for all time $t\in\mathbb{R}$ and $k=1,\ldots,K$. We call an oscillatory function satisfying these constraints an intrinsic mode type (IMT) function. Assume that the smallest frequency gap between two adjacent IMT components is $d$, and $d>0$, for all time $t\in\mathbb{R}$. That is, $\phi'_k(t)-\phi'_{k-1}(t)>d$, for $k=2,\ldots,K$. In practice, we assume $\epsilon<1$ and is small enough so that the AM is slowly varying. The function satisfying the above conditions is said to be the generalized AHM for the signal, and the constants $\epsilon,c,d$ are model parameters.

\subsection{STFT}

The STFT of a tempered distribution $x$ with respect to a chosen window $G$ in the Schwartz space is defined by
\begin{align}
{{V}}_{x}^{G}(u,\eta)= \int_{-\infty}^{\infty} x(t)G(t-u)e^{-i2\pi\eta(t-u)} \, \mathrm{d}t,  \label{STFT}
\end{align}
where $u\in\mathbb{R}$ is the time and $\eta\in\mathbb{R}^+$ is the frequency. 

\subsection{SST}

The SST can be embedded in different linear-type transforms, such as the CWT \cite{Daubechies_Lu_Wu:2011}, the wave packet \cite{Yang:2014} or the S-transform \cite{Huang_Zhang_Zhao_Sun:2015}.
Here we only mention the SST embedded in STFT due to the page limit.
The SST with the resolution $\kappa>0$ and the threshold $\gamma\geq 0$ is defined by
\begin{align}
{{S}}_{x}^{G,\kappa,\gamma}(u,\xi)=\int_{A_{x,\gamma}(u)} \hspace{-10pt}{{V}}_{x}^{G}(u,\eta)\frac{1}{\kappa}h\Big( \frac{|\xi-\omega^\gamma_x(u,\eta)|}{\kappa}   \Big) \, \mathrm{d}\eta ,  \label{S.0}
\end{align}
where $u\in\mathbb{R}$ is the time, $\xi>0$ is the frequency, $A_{x,\gamma}(u):=\left\{\eta\in\mathbb{R}_+:\left|{V}^{G}_x (u,\eta)\right|\geq\gamma \right\}$, $h(t)=\frac{1}{\sqrt\pi} e^{-t^2}$, $\kappa>0$ and $\omega_x(u,\eta)$ is the {\em reassignment rule}:
\begin{align}
\omega^\gamma_x (u,\eta)=\left\{
\begin{array}{ll}
\frac{-i\partial_u{{V}}_{x}^{G}(u,\eta)}{2\pi{{V}}_{x}^{G}(u,\eta)} &\mbox{when }|{V}_{x}^{G}(u,\eta)|\geq \gamma\\
-\infty & \mbox{when }|{{V}}_{x}^{G}(u,\eta)|<\gamma.
\end{array}
\right.	\label{S.0a}
\end{align}

The TFR determined by the STFT is sharpened by reassigning its coefficient at $(u,\eta)$ to a different point $(u,\xi)$ according to the reassignment rule. The SST is clearly nonlinear in nature.
It is important to note that the reassignment rule primarily depends on the phase information of the STFT, which contains the IF information.
According to the theoretical analysis in \cite{Wu:2011Thesis,Oberlin_Meignen_Perrier:2015}, the TFR of the SST is concentrated only on the IFs of all oscillatory components when the IF's of IMT functions in $x(t)$ are slowly varying.

While the SST algorithm looks complicated at the first glance, the idea underlying the algorithm is intuitive. Take a harmonic function $x(t)=Ae^{i2\pi \xi_0t}$ into account. Choose the window function $G$ that satisfies $\hat{G}$ is a real function and $\hat{G}(\xi)\geq\gamma$ when $\xi\in [-\Delta,\Delta]$, where $\gamma>0$ is chosen small enough and $\Delta>0$. Note that $x(t)$ is an IMT function. The STFT of $x(t)$ could be directly calculated by the Plancheral theorem, and we have
$V^{G}_x(u,\eta)=A\hat{G}(\eta-\xi_0)e^{i2\pi \xi_0 u}$. 
The information we have interest in an oscillatory signal, the IF, is hidden in the phase of $V^{G}_x(u,\eta)$. An intuitive idea to obtain the IF in this case is {first to apply the logarithm function on $V^{G}_x(u,\eta)$, next to divide it by $i2\pi$, and then to apply the derivative according to $u$ when $|\hat{G}(\eta-\xi_0)|\geq\gamma$; that is, $\frac{d}{i2\pi du}\big[\log(A\hat{G}(\eta-\xi_0))+i2\pi\xi_0 u\big]=\xi_0$.} Clearly, this operator is equivalent to the reassignment rule; that is, 
\begin{equation}
\partial_u\frac{\log[V^{G}_x(u,\eta)]}{i2\pi}=\frac{-i\partial_u{{V}}_{x}^{G}(u,\eta)}{2\pi{{V}}_{x}^{G}(u,\eta)}
\end{equation} 
when $|\hat{G}(\eta-\xi_0)|\geq\gamma$. We choose $\frac{-i\partial_u{{V}}_{x}^{G}(u,\eta)}{2\pi{{V}}_{x}^{G}(u,\eta)}$ to estimate the IF since we do not need to worry about the phase unwrapping problem when applying the logarithm function to a complex function.
To continue, note that we have $-i\partial_u V^G_x(u,\eta)=2\pi \xi_0V^{G}_x(u,\eta)$ by a direct calculation. Hence, $\omega^\gamma_x(u,\eta)=\xi_0$ when $\eta\in [\xi_0-\Delta,\xi_0+\Delta]$ and $\omega^\gamma_x(u,\eta)=-\infty$ otherwise. For this signal, we have $A_{x,\gamma}(u)=[\xi_0-\Delta,\xi_0+\Delta]$, and the reassignment rule indicates that the IF is $\xi_0$. Thus, the SST of $x$ can then be computed by the following equation:
\begin{align}
S^{G,\kappa,\gamma}_x(u,\xi)&=e^{i2\pi \xi_0 u} \int_{\xi_0-\Delta}^{\xi_0+\Delta} \hat{G}(\eta-\xi_0) \frac{1}{\kappa}\frac{1}{\sqrt{\pi}}e^{-|\xi-\xi_0|^2/\kappa^2}d\eta\\
&=Ce^{i2\pi \xi_0 u}\frac{1}{\kappa}e^{-|\xi-\xi_0|^2/\kappa^2}, \nonumber
\end{align}
where $C=\frac{1}{\sqrt{\pi}}\int_{-\Delta}^{\Delta} \hat{G}(\eta) d\eta\approx \frac{1}{\sqrt{\pi}}G(0)$. Clearly, when $\kappa$ is small, for each $u$, $S^{G,\kappa,\gamma}_x(u,\xi)$ is concentrated around $\xi_0$, which help alleviate the smearing effect in the STFT caused by the uncertainty principle.

\subsection{Second-Order SST}

When the IF is not slowly varying, the sharpening ability of the SST might be deteriorated. The 2nd-order SST resolves this problem by taking the second order information in the phase of the STFT to correct the reassignment rule. The 2nd-order SST could be viewed as a combination of the SST and the RM -- its sharpening ability is similar to that of the RM, and it allows us to reconstruct IMT functions like the SST. There are at least two versions of 2nd-order SST. We discuss the vertical SST (vSST) and the oblique SST (oSST) \cite{Oberlin_Meignen_Perrier:2015}.
Both the vSST and the oSST depend on the 2nd-order reassignment rule, which is a correction of the reassignment rule ${\omega}^\gamma_x$ in (\ref{S.0a}):
\begin{eqnarray}
\hat{\omega}^\gamma_x (u,\eta)=  
\left\{\begin{array}{ll}
 \stdp\omega^\gamma_x (u,\eta) + c(u,\eta)(u-\hat{t}_x(u,\eta)) &  \mbox{when }{\partial_{\eta} \hat{t}_x}(u,\eta)\ne0  \\

     {{\omega}^\gamma_x} (u,\eta) & \mbox{otherwise},
     \end{array}\right.	 \label{vsst.01}
\end{eqnarray}
where  $u\in\mathbb{R}$ is the time, $\eta>0$ is the frequency, and
\begin{eqnarray}
{{\hat t_x}(u,\eta) = u + i\frac{\partial_\eta{{V}}_{x}^{G}(u,\eta)}{{{V}}_{x}^{G}(u,\eta)}~ \mbox{and}~ c(u,\eta)=\frac{\partial_t{{\omega}^\gamma_x} (u,\eta)}{\partial_\eta \hat{t}_x (u,\eta)}.} \label{vsst.02}
\end{eqnarray}
The vSST with the resolution $\kappa>0$ and the threshold $\gamma\geq 0$ is defined by
\begin{align}
{{vS}}_{x}^{G,\kappa,\gamma}(u,\xi)=\int_{A_{x,\gamma}(u)} \hspace{-10pt}{{V}}_{x}^{G}(u,\eta)\frac{1}{\kappa}h\Big( \frac{|\xi-\hat{\omega}^\gamma_x(u,\eta)|}{\kappa}   \Big) \, \mathrm{d}\eta ; \label{vsst.03}
\end{align}
the oSST with the resolution $\kappa>0$ and $\tau>0$ and threshold $\gamma\geq 0$ is defined by
\begin{align}
{{oS}}_{x}^{G,\kappa,\gamma}(u,\xi)=\iint &{{V}}_{x}^{G}(y,\eta)e^{i\pi(2\xi-c(y,\eta)(\tau-y))(\tau-y)}\times\nonumber\\
&\frac{1}{\kappa}h\Big( \frac{|\xi-\hat{\omega}^\gamma_x(y,\eta)|}{\kappa}   \Big)\frac{1}{\tau}h\Big( \frac{|u-{\hat t_x}(y,\eta)|}{\tau}   \Big) \, \mathrm{d}\eta\mathrm{d}y.  \label{osst}
\end{align}
Note that the vSST could be viewed as a direct generalization of the SST with the modified reassignment rule, while the oSST could be viewed as a mixture of the SST and the RM. The reader is referred to \cite{Oberlin_Meignen_Perrier:2015} for details of the 2nd-order SST and \cite{Behera_Meignen_Oberlin:2015} for its theoretical analysis.

\subsection{IMT function reconstruction}
Each IMT function \cite{Daubechies_Lu_Wu:2011} can be reconstructed from the SST, as well as the vSST, if the input signal $x(t)=\sum_{k=1}^Kx_k(t)$ satisfies the AHM.
Take the SST as an example. Each IMT function $x_k=a_k(t)\cos(2\pi\phi_k(t))$, $k \in \{1,...,K\}$, can be reconstructed by the following two steps. First, evaluate the ``complexification'' of the $k$-th IMT function by
\begin{equation}\label{Formula:Reconstruction}
\hat{x}^{\mathbb{C}}_k(t) = \frac{1}{G(0)}\int_{\hat{\mathcal{Z}}_{k}(t)} {\tilde{S}}^{\kappa,\gamma}_{G,x}(t,\xi){\rm d}\xi ,
\end{equation}
where ${\hat{\mathcal{Z}}_{k}}(t)=[\hat{\phi}'_k(t)-\epsilon^{1/3},\hat{\phi}'_k(t)+\epsilon^{1/3}]$ and $\hat{\phi}'_k(t)$ is the estimated IF of the k-th IMT function, which can be obtained by the ridge extraction algorithm \cite{Chen_Cheng_Wu:2014,Carmona_Hwang_Torresani:1999,Meignen_Oberlin_McLaughlin:2012}. Then, the $k$-th IMT function is then extracted by 
\begin{equation}
\hat{x}_k(t)=\Re \hat{x}^{\mathbb{C}}_k(t),
\end{equation}
where $\Re$ is the operator taking the real part of the input complex value.
The reconstruction formula (\ref{Formula:Reconstruction}) could serve as an approach to obtain the complex form of a real signal. This property is important since, in general, evaluating the complex form of an IMT function is a nontrivial issue. It is opted that there are several constraints for the spectra of $a_k(t)$ and $\cos(2\pi\phi_k(t))$ in order to successfully obtain the imaginary counterpart of $x_k(t)$ and $a_k(t)\sin(2\pi\phi_k(t))$ via the Hilbert transform. We refer the reader with interest to \cite{Bedrosian:1962,Nuttall:1966} for details.

\section{Time-Varying Optimal Window Widths}\label{Section:TVOW}

It has been well-known that a short window is helpful for analyzing a signal with fast-varying IF components. On the other hand, for signals with two IMT functions with close IFs, the window should be long enough to avoid spectral overlaps. An ``optimal'' window should provide a balance between these two facts. 
However, the {uncertainty principle \cite{Grochenig:2001,Ricaud_Torresani:2014}} suggests that the benefits of a short and a long window width cannot be attained simultaneously.
In this regard, we need a method to choose a proper window width dynamically to balance on both ends. 

Several attempts have been proposed in the literature to balance between different window bandwidths. For example, in \cite{Jaillet_Torresani:2007,Ricaud:2014}, the TFJP was proposed to select the optimal window for the GT based on the R\'enyi entropy \cite{Coifman_Wickerhauser:1992}; in \cite{Balazs_Doerfler_Kowalski_Torresani:2013}, the NSGT depends on a frame associated with a non-uniform grid on the TF plane, which comes from the information provided by the signal. The frame could be viewed as the ``optimal window'' for the GT. These approaches have been shown to be helpful in the audio processing \cite{Jaillet_Torresani:2007}, for example, the beat tracking problem \cite{Holzapfel2011}. In general, these approaches could be understood as the TF tiling or a dictionary learning problem -- for a chosen redundancy, how to provide the best tiling of the TF plane, or to choose the optimal frame, so that the TF representation is ``optimal'' based on a chosen criterion, for example, the minimal $\ell^1$ norm \cite{Donoho_Elad:2003} or the minimal R\'enyi entropy. 

The reassignment-type transforms could be viewed as an approach to solve the dictionary learning problem by taking the phase of the STFT into account. Note that the STFT could be viewed as evaluating the coefficients of a signal associated with an infinitely redundant dictionary 
\begin{equation}\label{Definition:RedundantDictionary}
\mathcal{D}=\{G(t-\cdot)e^{i2\pi \xi t}\}_{t\in\mathbb{R},\xi\in\mathbb{R}^+}, 
\end{equation}
where $G$ is the chosen window. Directly determining the optimal frame out of $\mathcal{D}$ is not an easy task. Instead of determining the optimal frame, the reassignment rule used in the RM and the SST and its variations could be viewed as an alternative to approximate the optimal frame out of $\mathcal{D}$. 
Note that in the SST (\ref{S.0}), the vSST (\ref{vsst.03}), and the oSST (\ref{osst}), the coefficients of the STFT are moved to a new location based on the reassignment rule. 
In this sense, nonlinear-type TF analysis could be viewed as evaluating the coefficients of an approximated optimal frame. We mention that this viewpoint has been taken into account to design the Tycoon algorithm \cite{Kowalski_Meynard_Wu:2015}. Theoretically, if the signal satisfies the AHM model, it has been shown that the reassignment rule could lead to the optimal frame \cite{Chen_Cheng_Wu:2014,Daubechies_Lu_Wu:2011,Oberlin_Meignen_Perrier:2015}. However, due to the lack of knowledge of the model parameters, like $\epsilon,c,d$ of a given signal, the reassignment rule, and hence the TFR, might be influenced by the interaction of the chosen window and the time-varying AM and IF, and the overlap of spectra of different oscillatory components. In practice, although we have a rule of thumb of how to choose the window based on the a priori knowledge of the signal, the reassignment rule might deviate from the optimal frame.

In order to resolve this issue, we propose an adaptive way to determine the optimal window for the reassignment-type transforms. This approach can be viewed as correcting the approximated optimal frame determined by the reassignment rule. 
A window is regarded as optimal for a chosen reassignment-type TF analysis if it provides the most concentrated TFR. Since the IF and AM of each IMT function may vary from time to time, a single window optimal for the entire signal might not be suitable. Therefore, the notion of the optimal window for a chosen TF analysis should be \textit{local}. For example, for each time, we determine an optimal window.
In general, finding the optimal window is a difficult task. In statistics, the problem is commonly reduced to the \textit{window bandwidth selection} problem \cite{Wand_Jones:1995}. In this work, we simplify the window selection problem to the window bandwidth selection problem.
To further simplify the discussion, we consider the Gaussian window, that is, 
\begin{equation}\label{Definition:GaussianWindow}
G(t)={g_\sigma}(t) := \frac{1}{\sqrt{2\pi}\sigma} e^{-t^2/(2\sigma^2)}, 
\end{equation}
where $\sigma>0$ is the \textit{bandwidth} of the window. In this case, the STFT is the same as the GT. 
In this section, for a chosen TF analysis with the Gaussian window (\ref{Definition:GaussianWindow}), we describe a time-varying optimal window width (TVOWW) selection scheme and an adaptive optimal window width (AOWW) selection scheme to compute a series of local optimal window widths. 
We mention that although we focus on the window bandwidth selection problem with the Gaussian window, the discussion below could be directly generalized to other window functions, or even multiple window functions.

\subsection{The TVOWW and the AOWW selection schemes}

First select a reassignment-type transform, for example, the SST.
The TVOWW selection scheme evaluates the local window width by iterating the following steps for each time $u\in\mathbb{R}$:

\begin{enumerate}

\item Evaluate the distribution concentration of the TFR on $[u-b,u+b]\times \mathbb{R}^+$, where $b\geq 0$ determines the size of the neighborhood, by a chosen distribution concentration measure, denoted as $\mathrm C_{\sigma,b}(u)$.

\item The local optimal window width at the time instant $u$ is determined by
\begin{equation}
\tilde{\sigma}_b(u):=\text{argmin}_{\sigma>0}\mathrm C_{\sigma,b}(u).   \label{VOW.3}
\end{equation}

\item Apply the window width $\tilde{\sigma}_b(u)$ to evaluate the TFR of the signal $x(t)$ at time $u$. 
\end{enumerate}

The proposed scheme could be directly applied to other TF analyses, such as the STFT, the 2nd-order SST or other nonlinear TF analyses. When $b=\infty$, $\tilde{\sigma}_b$ is a constant value and the SST is reduced to the original SST with one window width, which is chosen to optimize the selected measure of distribution concentration. We regard this special case the \textit{global optimal window width} (GOWW).

The AOWW selection scheme evaluates the local window width via iterating the following steps for {a given pair} of time and frequency, $(u,\xi)$. 
\begin{enumerate}

\item Evaluate the distribution concentration of the TFR on $[u-b,u+b]\times[\max\{0,\xi-b_F\},\xi+b_F]$, where $b\geq 0$ determines the size of the neighborhood and $b_F>0$ determines the size of the neighborhood in the frequency axis, by a chosen distribution concentration measure, denoted as $\mathrm C_{\sigma,b,b_F}(u,\xi)$.

\item The local optimal window width at the time instant $u$ is determined by
\begin{equation}
\tilde{\sigma}_{b,b_F}(u,\xi):=\text{argmin}_{\sigma>0}\mathrm C_{\sigma,b,b_F}(u,\xi).   \label{VOW.4}
\end{equation}

\item Apply the window width $\tilde{\sigma}_{b,b_F}(u,\xi)$ to evaluate the TFR of the signal $x(t)$ at time $u$ and frequency $\xi$. 
\end{enumerate}

While the AOWW could provide a sharper TFR, compared with the TVOWW, the computational burden of the AOWW selection scheme is greatly increased. Furthermore, for a given time $u$, since the window width varies for different frequencies, the reconstruction formula (\ref{Formula:Reconstruction}) cannot be applied. 
{We mention that the above algorithm can be easily generalized to selecting multiple window functions. Thereby, different windows can be taken into account in the optimization (\ref{VOW.3}) or (\ref{VOW.4}), so that the optimal window function and its corresponding optimal window width are selected. Since the multiple window selection is out of the scope of this work, we will study it in the future work.}

\subsection{R\'enyi entropy}

The information entropy is a common measure to estimate the dispersion of an information content.
By viewing the TFR at each time as a probability density function, a larger entropy indicates a less distributed concentration of the TFR. In this study, we adopt the R\'enyi entropy to measure the distribution concentration of a TFR \cite{Baraniuk_Flandrin_Janssen_Michel:2001}.

The $\alpha$-R\'enyi entropy of a non-zero function $p$, where $\alpha>0$, is defined as
\begin{align}
 {R}_{\alpha}(p):=\frac{1}{1-\alpha} \log_2 \left(\frac{\|p\|_{2\alpha}}{\|p\|_2}\right)^{2\alpha},
\end{align}
{where $\|p\|_\alpha:=(\int |p(x)|^\alpha d x)^{1/\alpha}$ for $0<\alpha<\infty$. Note that when $\alpha<1$, $\|\cdot\|_\alpha$ is not a norm but a quasi-norm.}
It is well-known that the larger the R\'enyi entropy is, the less concentrated the distribution is \cite{Jaillet_Torresani:2007,Stankovic:2001}. 
That is to say, a window width providing the least R\'enyi entropy is regarded as the optimal window width.
Note that when $\alpha\to 0$, the R\'enyi entropy gives the $\ell^0$ norm information of the signal; when $\alpha\to 1$, the Shannon entropy is recovered; and when $\alpha\to 1/2$, we obtain the information of the commonly used ratio norm $\ell^1/\ell^2$. 
In general, $\alpha>2$ is recommended for TFR measures \cite{Stankovic:2001} and we chose $\alpha=2.4$ in this study.
{In practice, we notice that the results are insensitive within a certain range of $\alpha$ values ($\alpha>0$).}

Denote the TFR of a chosen TF analysis $P$ defined on $\mathbb{R}\times \mathbb{R}^+$. The TFR distribution is considered the most concentrated if its corresponding R{\'e}nyi entropy is minimized. We thus define the measure of distribution concentration in the TVOWW selection scheme as  
\begin{align}
 C_{\sigma,b}(u)\label{RenyiDefinition}:=\frac{1}{1-\alpha} \log_2\frac{\iint_{I_{u}} |R(t,\xi)|^{2\alpha}\mathrm {d}t\mathrm {d}\xi}{\big(\iint_{I_{u}} |R(t,\xi)|^2 \mathrm {d}t\mathrm {d}\xi\big)^\alpha},
\end{align}
where $u\in\mathbb{R}$ and $I_u:=[u-b,u+b]\times [0,\infty)$. Similarly, the distribution concentration measure in the AOWW selection scheme is defined as
\begin{align}
C_{\sigma,b,b_F}(u,\xi)\label{RenyiDefinition2}
:=\frac{1}{1-\alpha} \log_2\frac{\iint_{J_{u,\xi}} |R(t,\xi)|^{2\alpha}\mathrm {d}t\mathrm {d}\xi}{\big(\iint_{J_{u,\xi}} |R(t,\xi)|^2 \mathrm {d}t\mathrm {d}\xi\big)^\alpha}, 
\end{align}
where $u\in\mathbb{R}$, $\xi\in\mathbb{R}^+$, and $J_{u,\xi}:=[u-b,u+b]\times[\max\{0,\xi-b_F\},\xi+b_F]$.

\section{Results and Discussions}\label{Section:Results}

We start the demonstration of the proposed the TVOWW and the AOWW selection schemes by analyzing a synthetic data. We then show the result of analyzing the laser-driven atomic dipole moment, and discuss the performance of the proposed scheme. 
In this section, for the SST and the 2nd-order SST, the numerical value of $\kappa$ and $\tau$ are selected to be small enough so that $\frac{1}{\kappa}h(\frac{\cdot}{\kappa})$ and $\frac{1}{\tau}h(\frac{\cdot}{\tau})$ are both implemented as discretized Dirac measures. The $\gamma$ value is fixed at $10^{-6}\%$ of the mean square energy of the signal $x(t)$ under analysis.

\subsection{Synthetic Signal} \label{SubSection:Synthetic}
Consider a multicomponent signal given by
%
%
\begin{align}
     &x(t) = x_1(t)+x_2(t)+x_3(t),          \label{Synthetic}
\end{align}
where the signal components are:
\begin{align}
&x_1(t) = \cos(2\pi\phi_1(t))\chi_{[-\infty, 20]}(t)  \nonumber \\ 
&x_2(t) = \cos(2\pi\phi_2(t))\chi_{[-\infty, 13.6]}(t) \nonumber \\ 
&x_3(t) = \cos(2\pi\phi_3(t))\chi_{[17.5,\infty]}(t), \nonumber
\end{align}
where $\chi_I$ is the indicator function supported on $I\subset \mathbb{R}$ and 
\begin{align}
& \phi_1(t)=1.33^{t-5}+3t \nonumber \\
& \phi_2(t)=-0.0437(t-5)^4+0.5(t-5)^3+0.25(t-5)^2+5t  \nonumber \\
& \phi_3(t)=-\frac{2.7}{3.5}\cos{(3.5t)+0.85(t-15)^2+0.5t}.  \nonumber
\end{align}
The corresponding IFs are $\phi'_1(t)=({\ln 1.33})1.33^{t-5}+3$, $\phi'_2(t)=-0.175(t-5)^3+1.5(t-5)^2+0.5(t-5)+5$, and $\phi'_3(t)=2.7{\sin 3.5t}+1.7(t-15)+0.5$.
The observed signal $Y(t)=x(t)+\lambda\Phi(t)$, where $\Phi$ is the white Gaussian noise with mean $0$ and standard deviation (std) $1$,  and the $\lambda$ value ($\lambda>0$) is chosen so that the signal-to-noise ratio (SNR), defined as $20\log\frac{\text{std}(x(t))}{\lambda}$, is $15$ dB. $Y(t)$ is sampled at $60$ Hz from the $0$-th to the $25$-th second (s). We select the optimal window width $\sigma$ from a set of candidate bandwidths, $\{11/720,31/720\ldots,501/720\}$ s.

\subsubsection{TFR with the GOWW}

We first show the limitation of using the GOWW selection scheme for the SST. In other words, we run the optimal window selection scheme with $b=\infty$, resulting in the GOWW of $71/720$ s. 
Fig.~\ref{TVOW}(a) demonstrates that the SST with the GOWW can capture the oscillatory dynamics.
Nevertheless, while a small window width is required to reduce the R{\'e}nyi entropy in the TFR, it results in the evident interference pattern between the neighboring IF components.
For example, the spectral gap (differences between the adjacent IF components) at the $5$-th s (i.e.,$\phi_1'(5)$ and $\phi_2'(5)$) is $1.8$ Hz, and a strong interference pattern is observed at the $5$-th s.
According to Definitions 3.1 and 3.2 in \cite{Daubechies_Lu_Wu:2011}, the window width, measured by the full width at half maximum (FWHM), which is defined as $2\sqrt{2\ln2}{\tilde\sigma_b}$, should be at least $1/1.8\approx 0.55$ s in order to separate the two neighboring components in the AHM.
Here, the FWHM of the GOWW is $0.23$ s, which is insufficient and leads to the interference pattern.
Similar interference patterns can be observed at times $13.6$ s, and $18.5$ s, where spectral gaps are approximately 3 Hz and 7 Hz, respectively.
It is clear that a larger spectral gap results in a less coupled interaction between the IF components.
In summary, since the optimal window is chosen globally, the local details may not be refined even if the overall sharpness of the TFR is increased.
%


\subsubsection{TFR with the TVOWW}

\begin{figure}[h]
\begin{subfigure}{0.45\textwidth}
		\includegraphics[width=\linewidth]{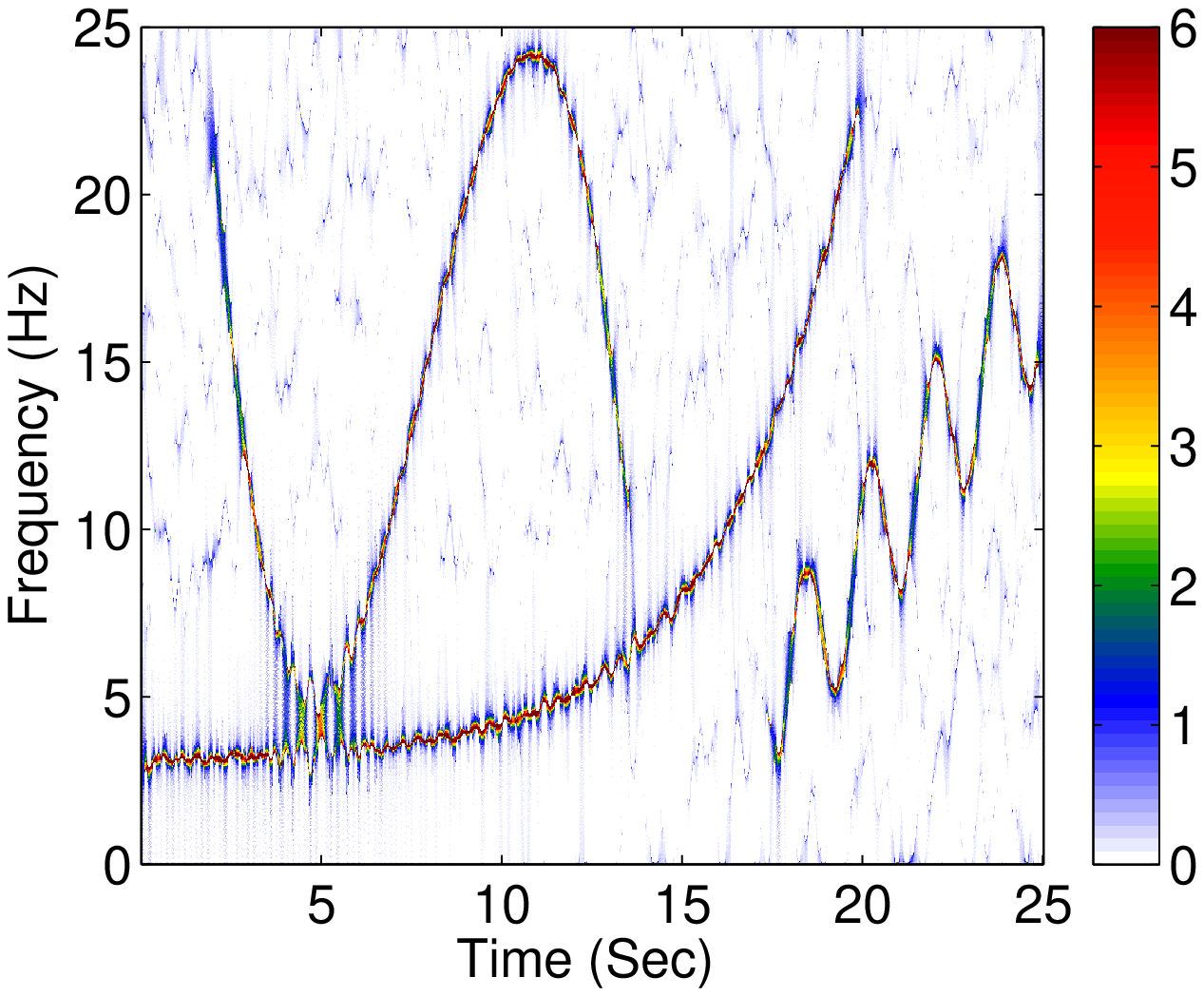}
\caption{SST with the GOWW} \label{TVOW:a}
\end{subfigure}
\begin{subfigure}{0.45\textwidth}
		\includegraphics[width=\linewidth]{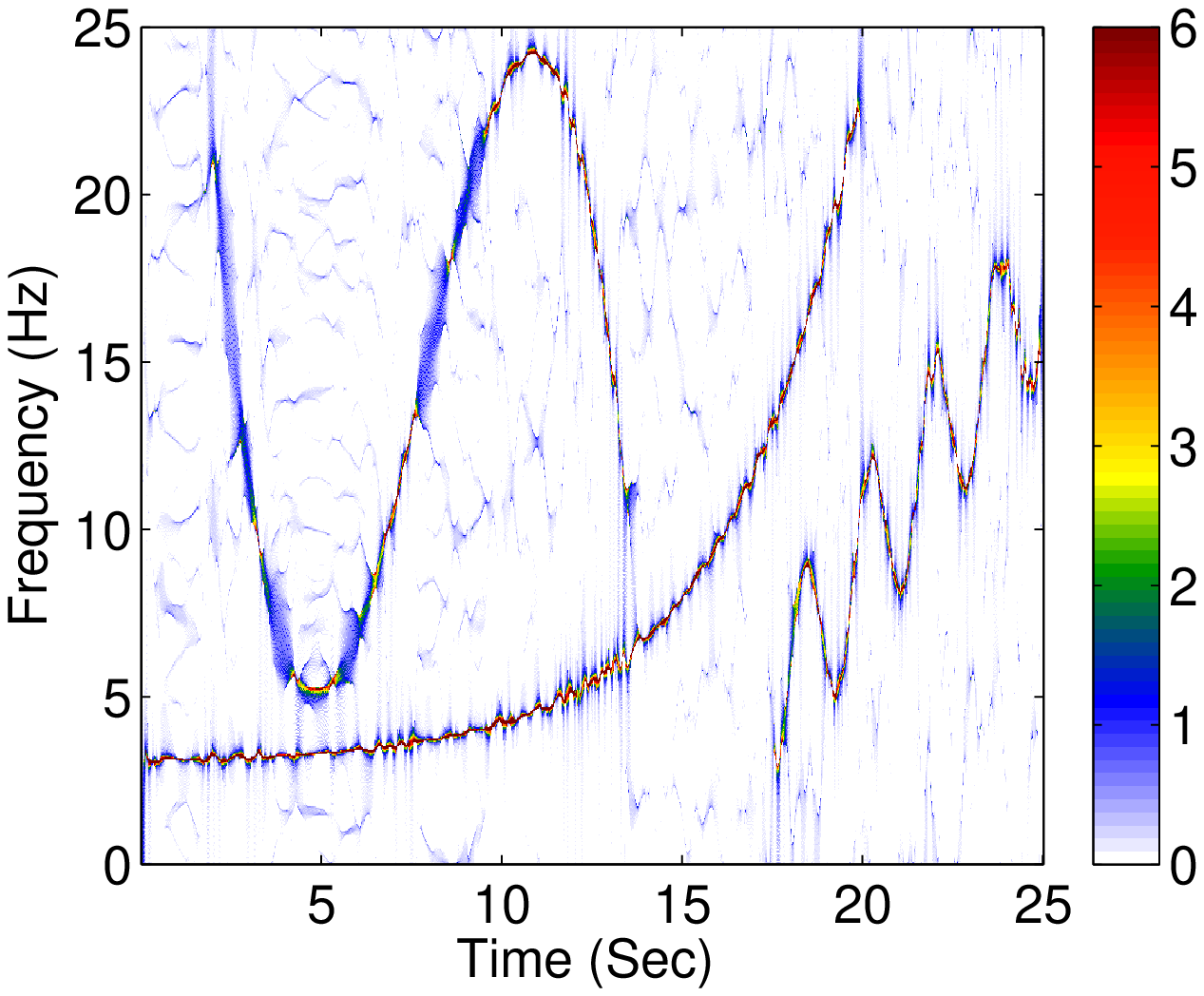}
\caption{SST with TVOWW} \label{TVOW:b}
\end{subfigure}\\
\begin{subfigure}{0.45\textwidth}
		\includegraphics[width=\linewidth]{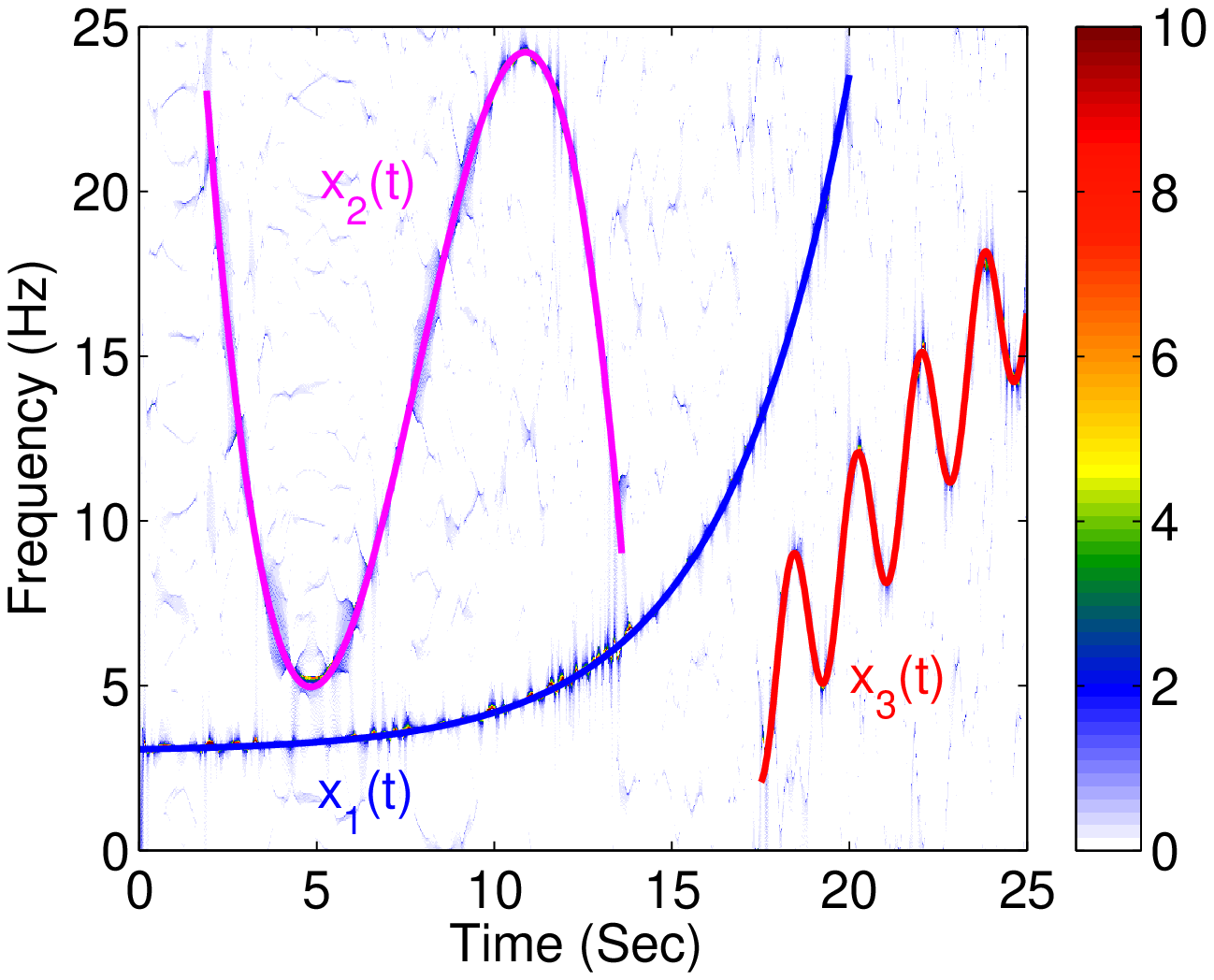}
\caption{Comparison with ideal IFs} \label{TVOW:c}
\end{subfigure}
\begin{subfigure}{0.45\textwidth}
		\includegraphics[width=\linewidth]{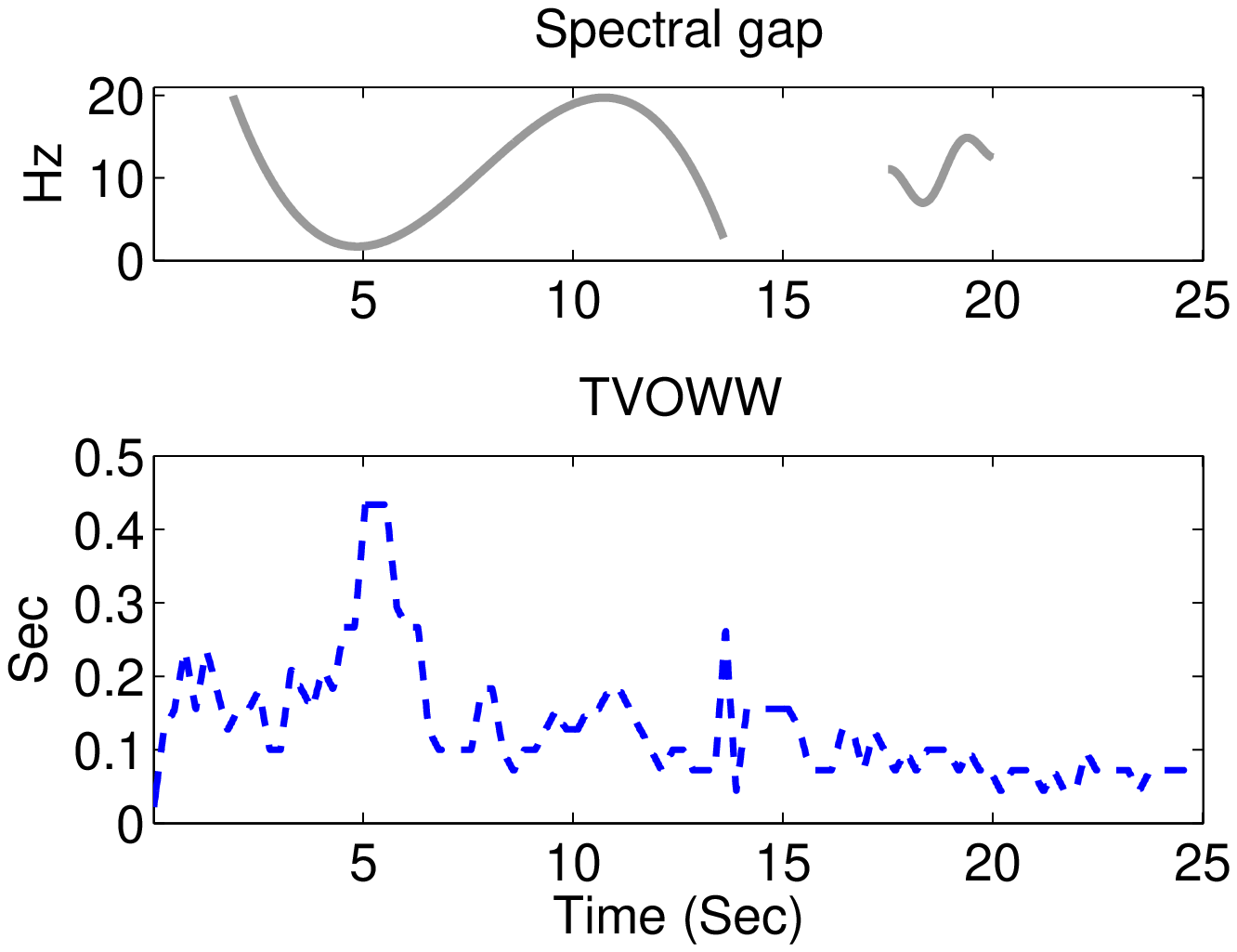}\\
\caption{Spectral gap and the TVOWW} \label{TVOW:d}
\end{subfigure}
\caption{(a) The TFR of the SST with the GOWW. (b) The TFR of the SST with the TVOWW. (c) The true IFs (blue:$x_1(t)$; magenta:$x_2(t)$; red:$x_3(t)$) are superimposed on the TFR of the SST with the TVOWW. Note that the range of the colorbar is increased for the comparison. (d) The spectral gap (upper panel) and the corresponding TVOWW (lower panel). 
It is clear that when the spectral gap is small, a longer window is needed. The TFR values are normalized by the z-score.
 }
  \label{TVOW}
\end{figure}

We next demonstrate that the proposed TVOWW selection scheme can further improve the TFR quality.
To reduce the computation capacity, we evaluate the local optimal window width every $0.25$ s in a neighborhood with a width of $2b=0.33$ s. 
The neighborhood size is found to be insensitive to the final result.
While a small value is favorable, the width of the neighborhood should be greater than the sampling period \cite{Hlawatsch_Flandrin:1997}.
Subsequently, a linear interpolation is applied to the samples of the TVOWWs such that there is an optimal window for each time instant in the signal interval.
The TFR of the SST with the TVOWW is presented in Fig.~\ref{TVOW}(b) and its comparison with the true IFs is displayed in Fig.~\ref{TVOW}(c).
It is clearly shown that the coupling artifact between closing IF components is eliminated, particularly at the $5$th s, as well as at the $13.6$, and $18.5$ s.
The IF components in the TFR with an improved quality approaches the ideal IF components, as in Fig.~\ref{TVOW}(c).
We further display the corresponding TVOWW along with the spectral gap in Fig.~\ref{TVOW}(d).
According to this figure, the window widths become large at the closing times $5$ s, $13.6$ s, and $18.5$ s to separate the different IF components.
Note that at time $5$ s, the largest window width is ${\tilde\sigma_b}=\frac{345}{720}=0.48$ s, corresponding to a FWHM of $1.08$ s, which is larger than $0.55$ s.

\subsubsection{Necessity of Selecting a Proper Window Width}

\begin{figure}[h]
\centering

\begin{subfigure}{0.42\textwidth}
		\includegraphics[width=\linewidth]{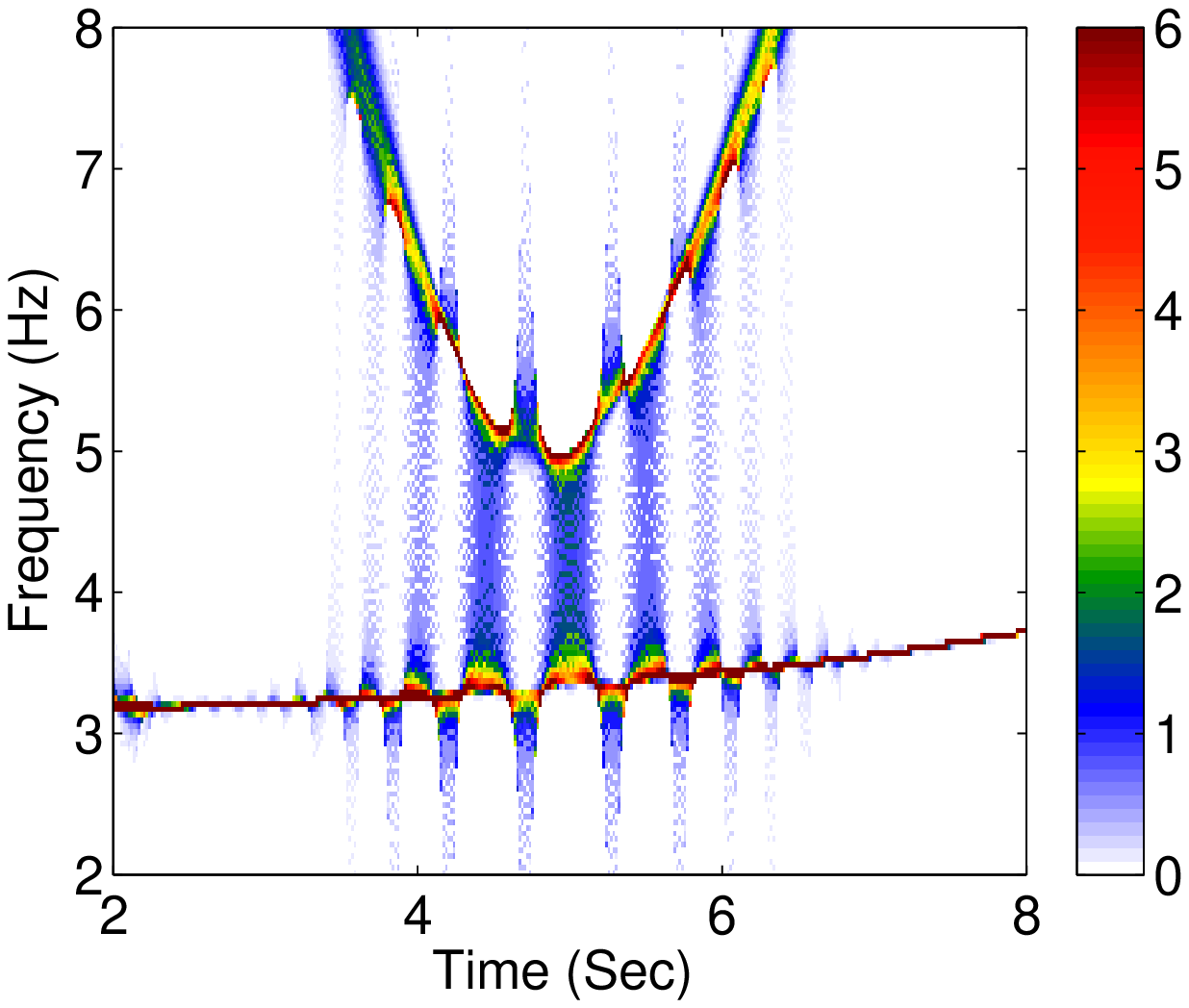}
\caption{SST, small window width} \label{Coupling:a}
\end{subfigure}
\begin{subfigure}{0.42\textwidth}
		\includegraphics[width=\linewidth]{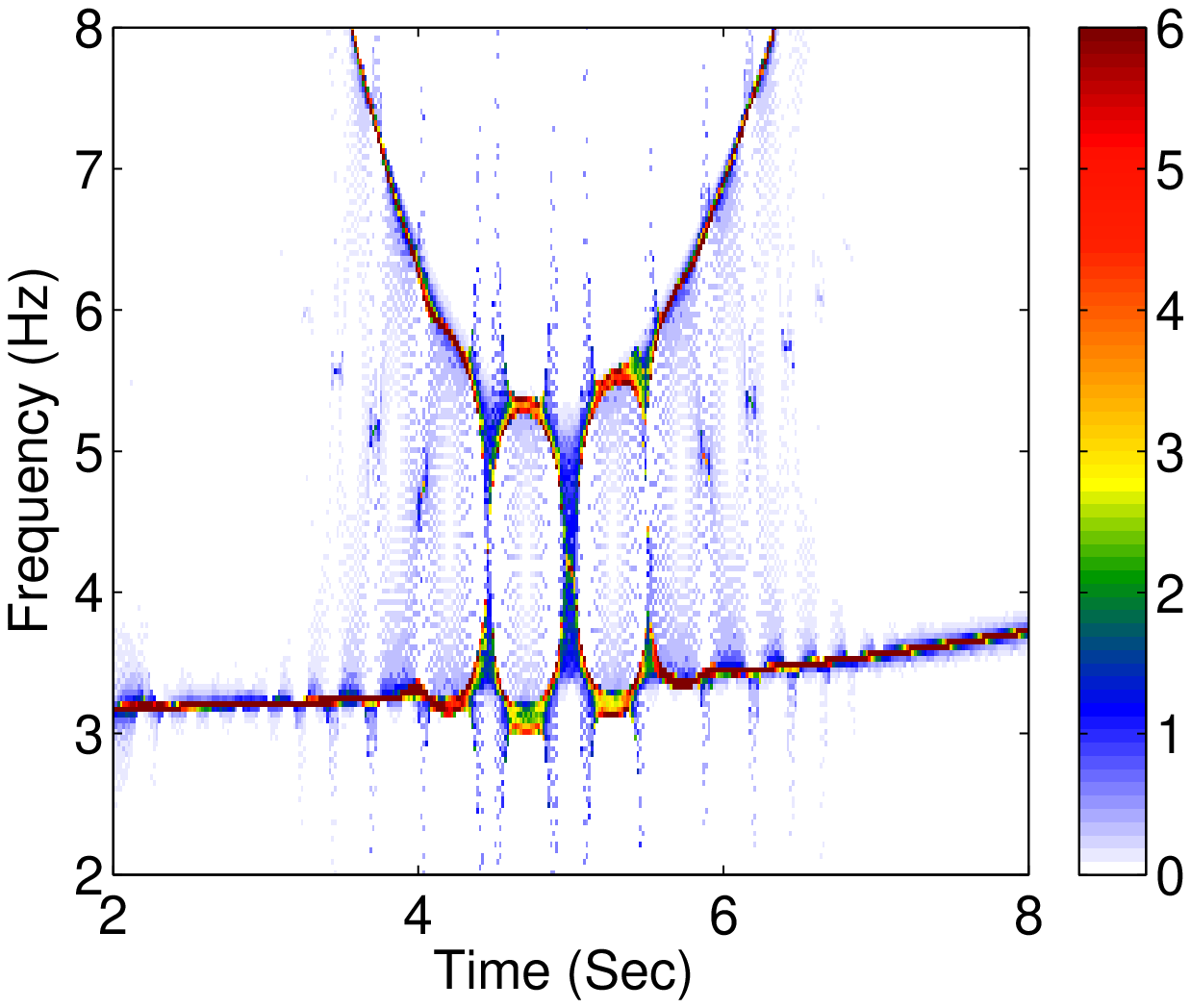}
\caption{vSST, small window width} \label{Coupling:b}
\end{subfigure}
\begin{subfigure}{0.42\textwidth}
		\includegraphics[width=\linewidth]{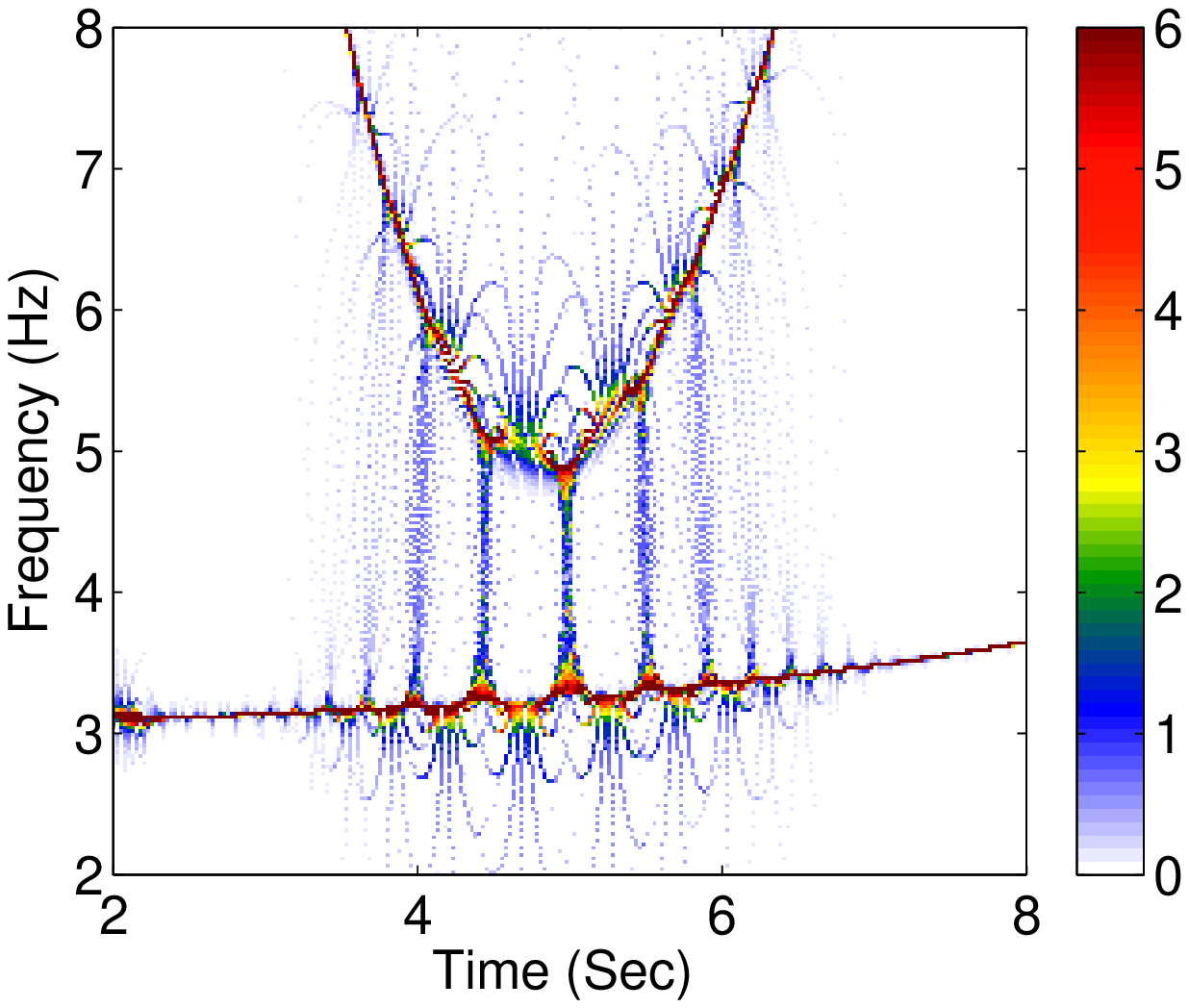}
\caption{oSST, small window width} \label{Coupling:c}
\end{subfigure}
\begin{subfigure}{0.42\textwidth}
		\includegraphics[width=\linewidth]{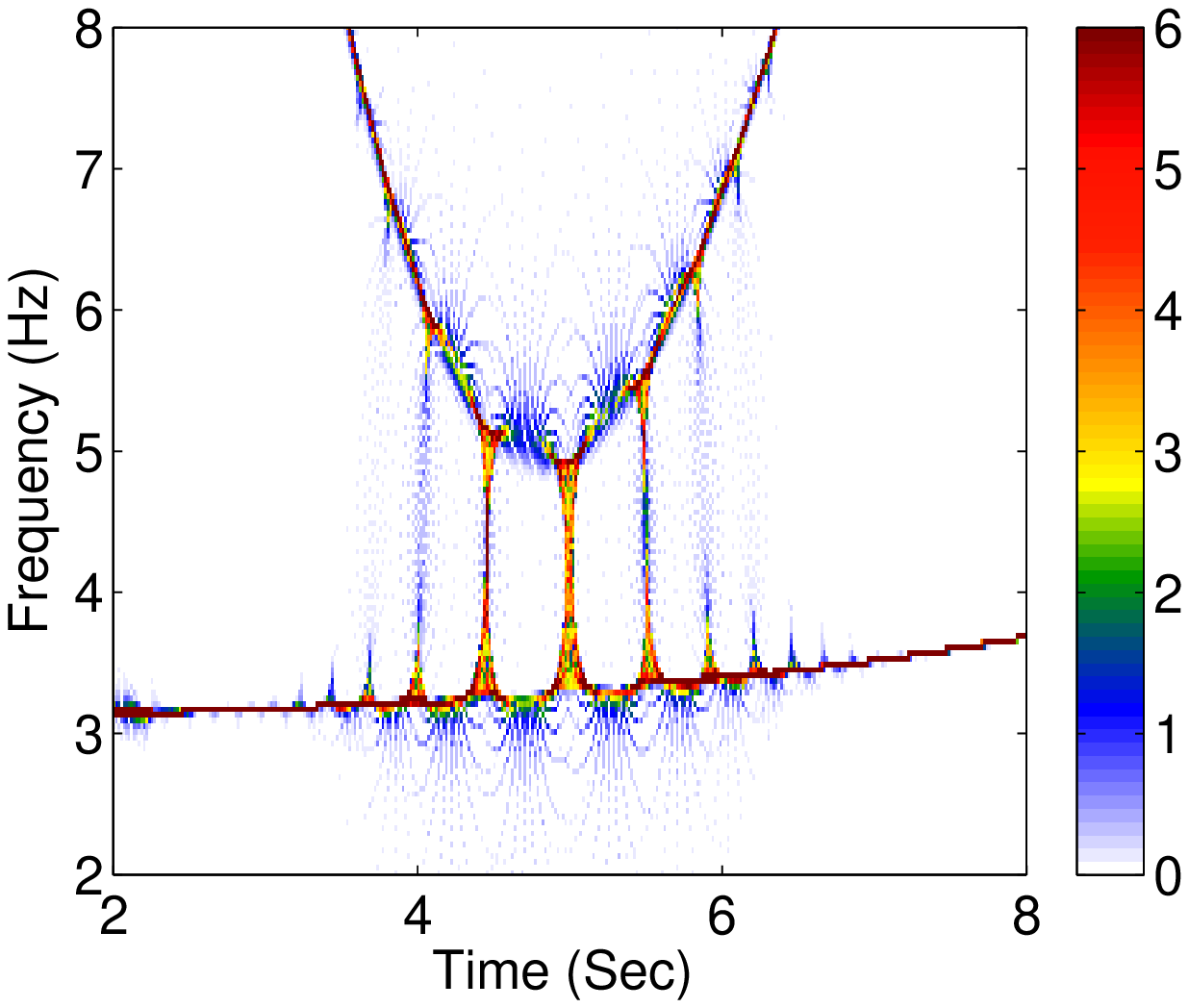}
\caption{RM, small window width} \label{Coupling:d}
\end{subfigure}

\begin{subfigure}{0.42\textwidth}
		\includegraphics[width=\linewidth]{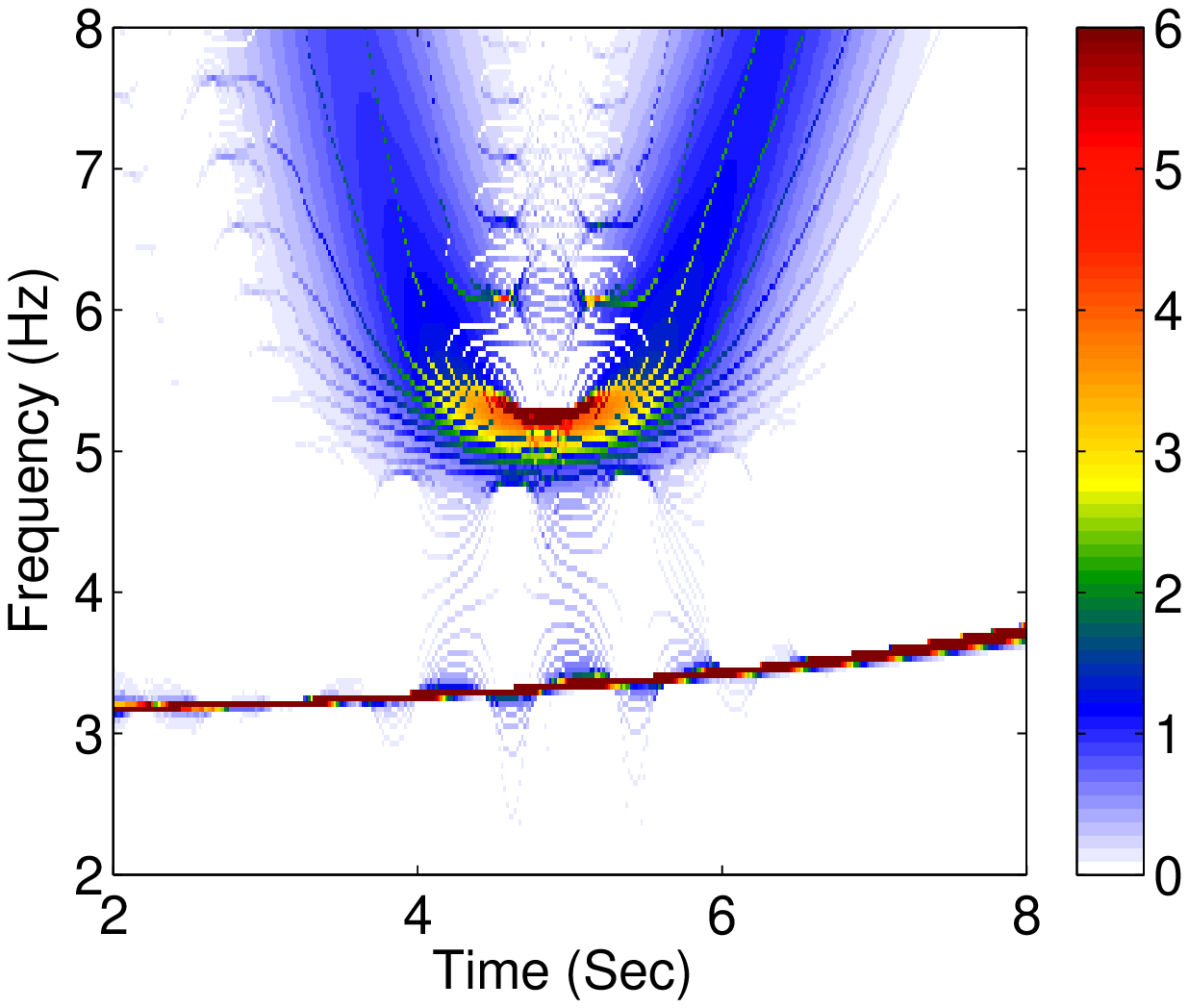}
\caption{SST, large window width} \label{Coupling:e}
\end{subfigure}
\begin{subfigure}{0.42\textwidth}
		\includegraphics[width=\linewidth]{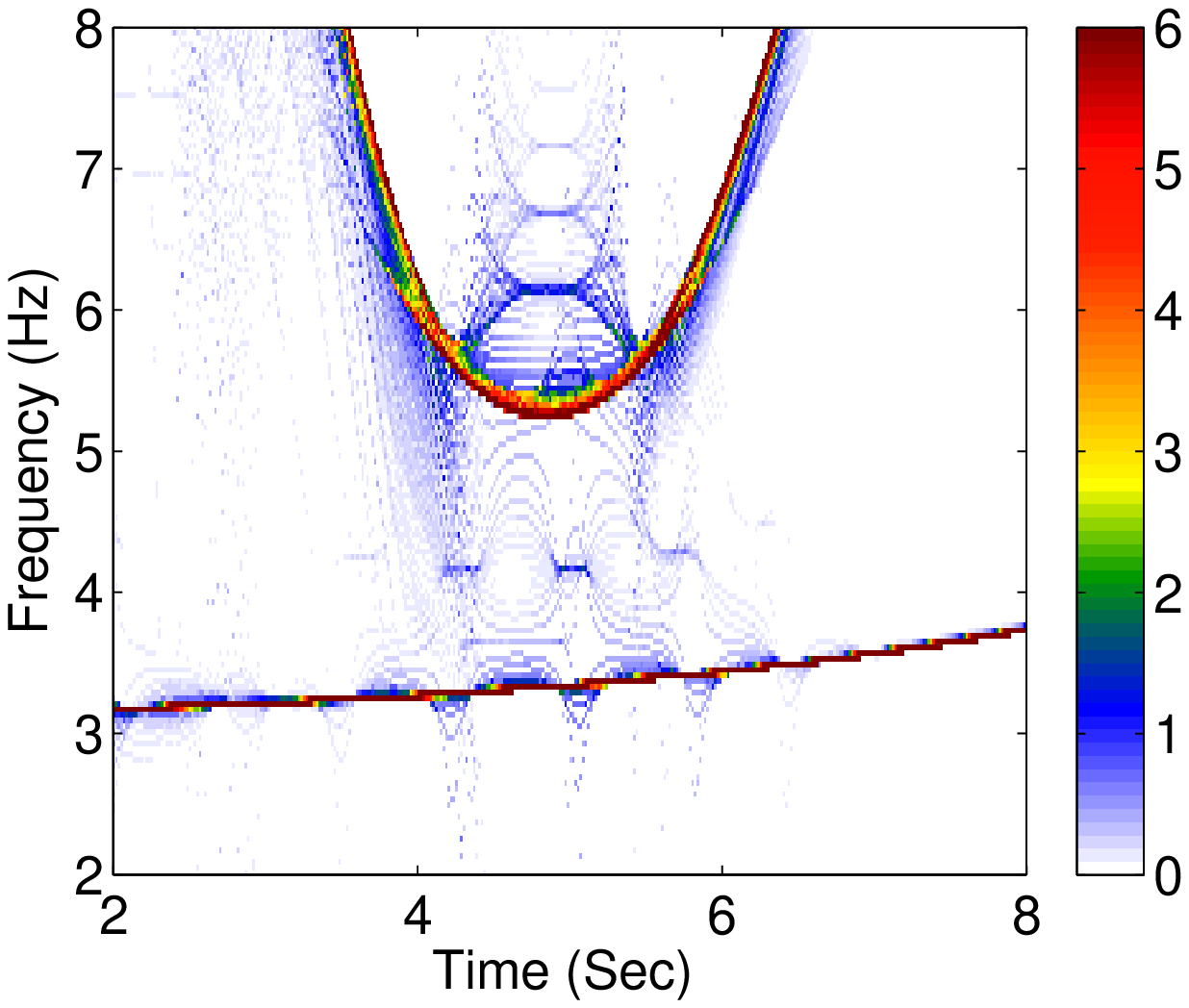}
\caption{vSST, large window width} \label{Coupling:f}
\end{subfigure}
\begin{subfigure}{0.42\textwidth}
		\includegraphics[width=\linewidth]{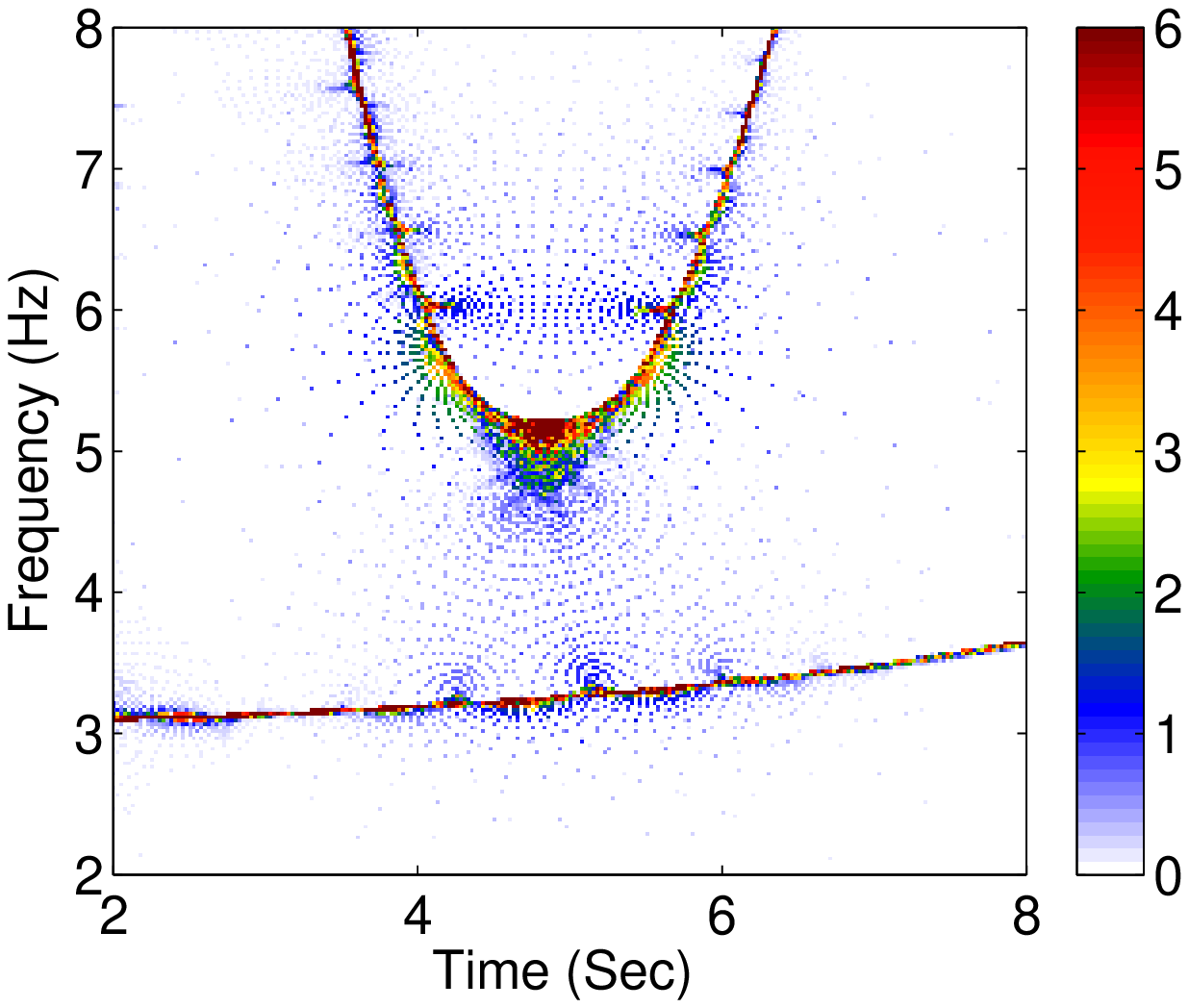}
\caption{oSST, large window width} \label{Coupling:g}
\end{subfigure}
\begin{subfigure}{0.42\textwidth}
		\includegraphics[width=\linewidth]{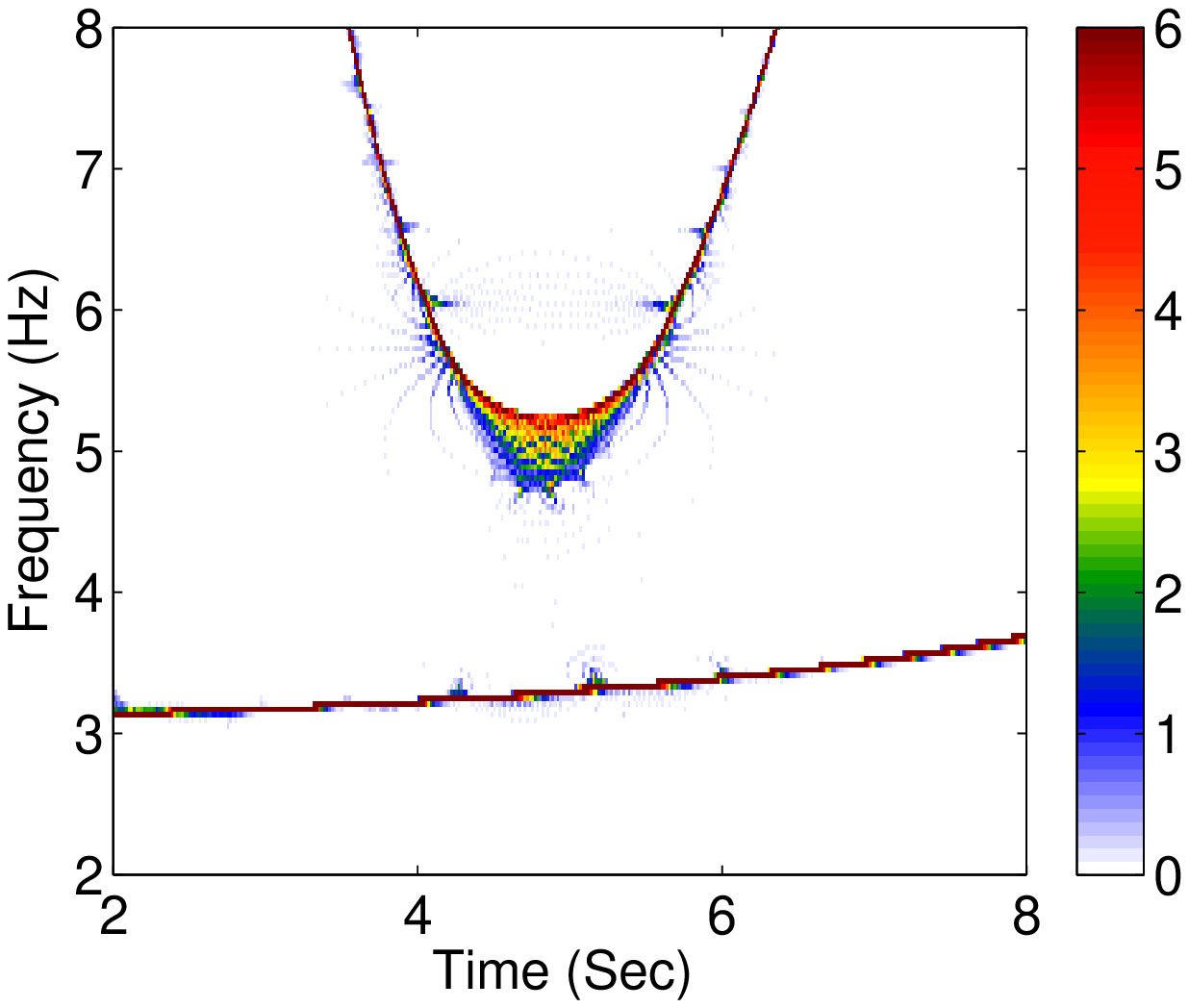}
\caption{RM, large window width} \label{Coupling:h}
\end{subfigure}

\caption{\label{Coupling} A large window is needed to separate the two adjacent IF components for different TF analyses.
The window width in the upper panel is $\frac{111}{720}$ s (small), and that in the lower panel is $\frac{345}{720}$ s (large).
The TFRs of the SST, the vSST, the oSST, and the RM are shown in (a)(e), (b)(f), (c)(g), and (d)(h), respectively.
The TFR values are normalized by the z-score.
It is clear that while the TFRs of the 2nd-order SST and the RM are sharpened, with the small window width these TFRs suffer from the ``coupling artifact'' caused by the two closing IF components. A longer window width in this case can help remove the artifact. We could see that the SST could not well handle the fast-varying IF.}
\end{figure}

\begin{figure}[h]
\centering

\begin{subfigure}{0.42\textwidth}
		\includegraphics[width=\linewidth]{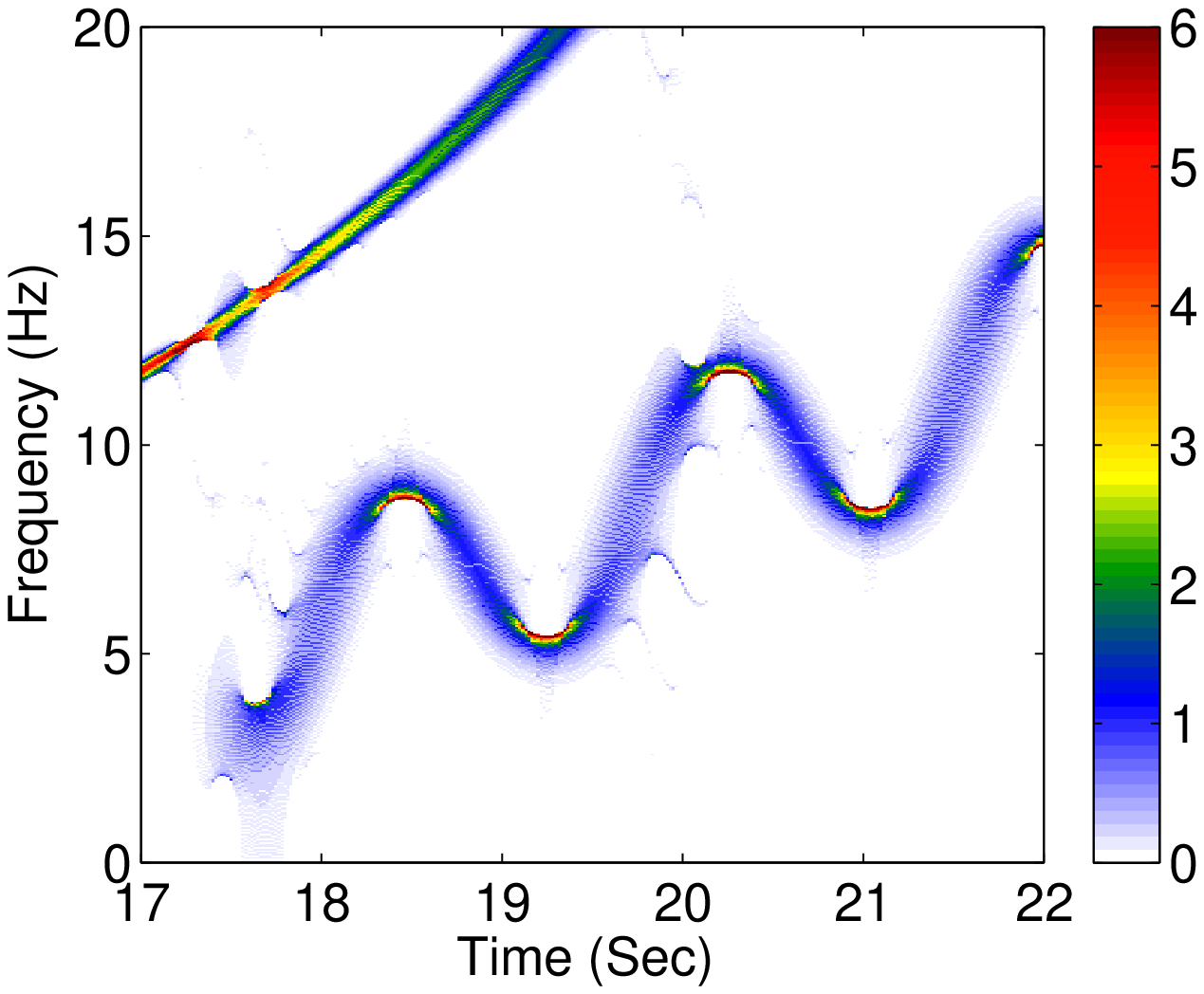}
\caption{SST, small window width} \label{Symmetry:a}
\end{subfigure}
\begin{subfigure}{0.42\textwidth}
		\includegraphics[width=\linewidth]{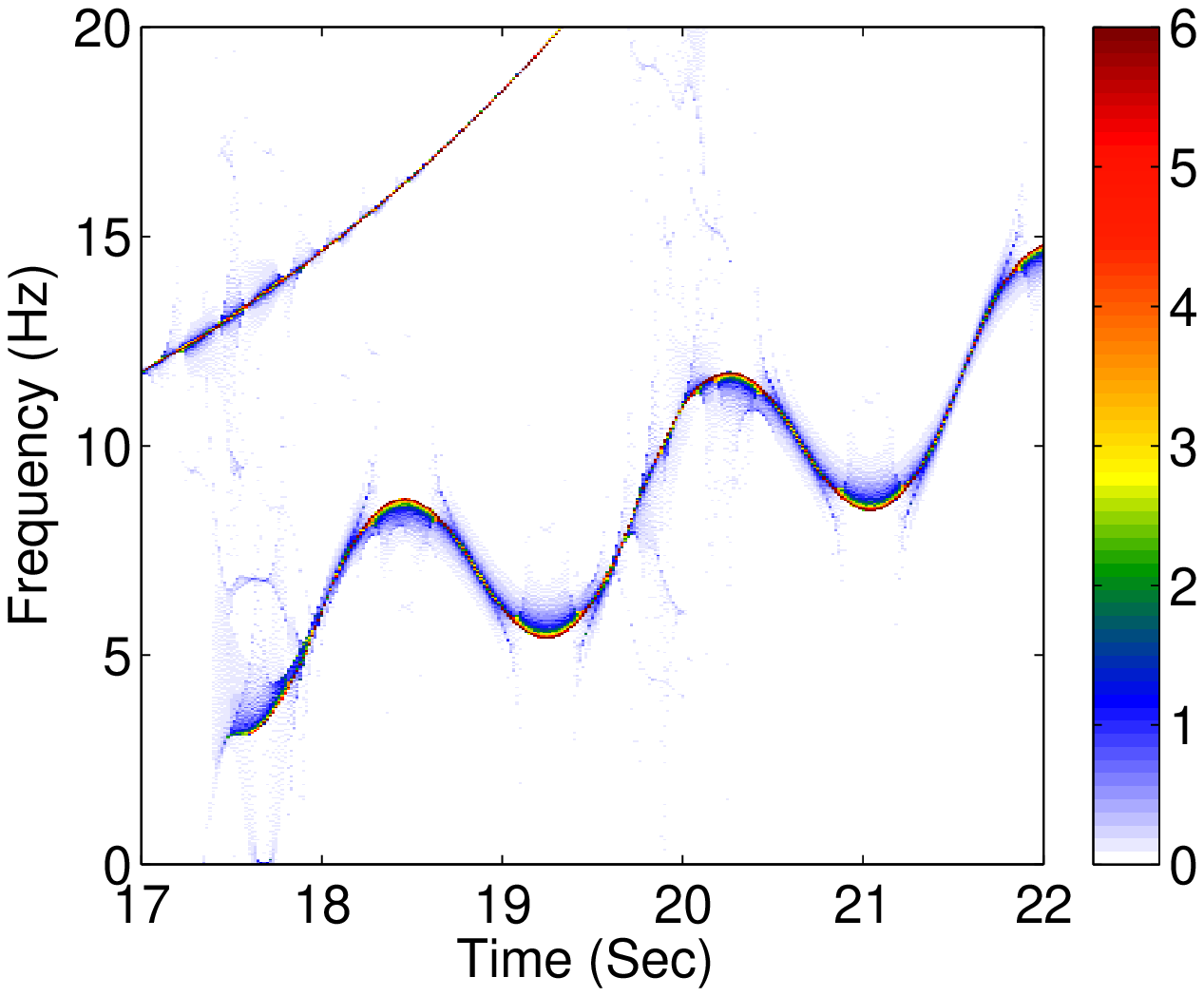}
\caption{vSST, small window width} \label{Symmetry:b}
\end{subfigure}
\begin{subfigure}{0.42\textwidth}
		\includegraphics[width=\linewidth]{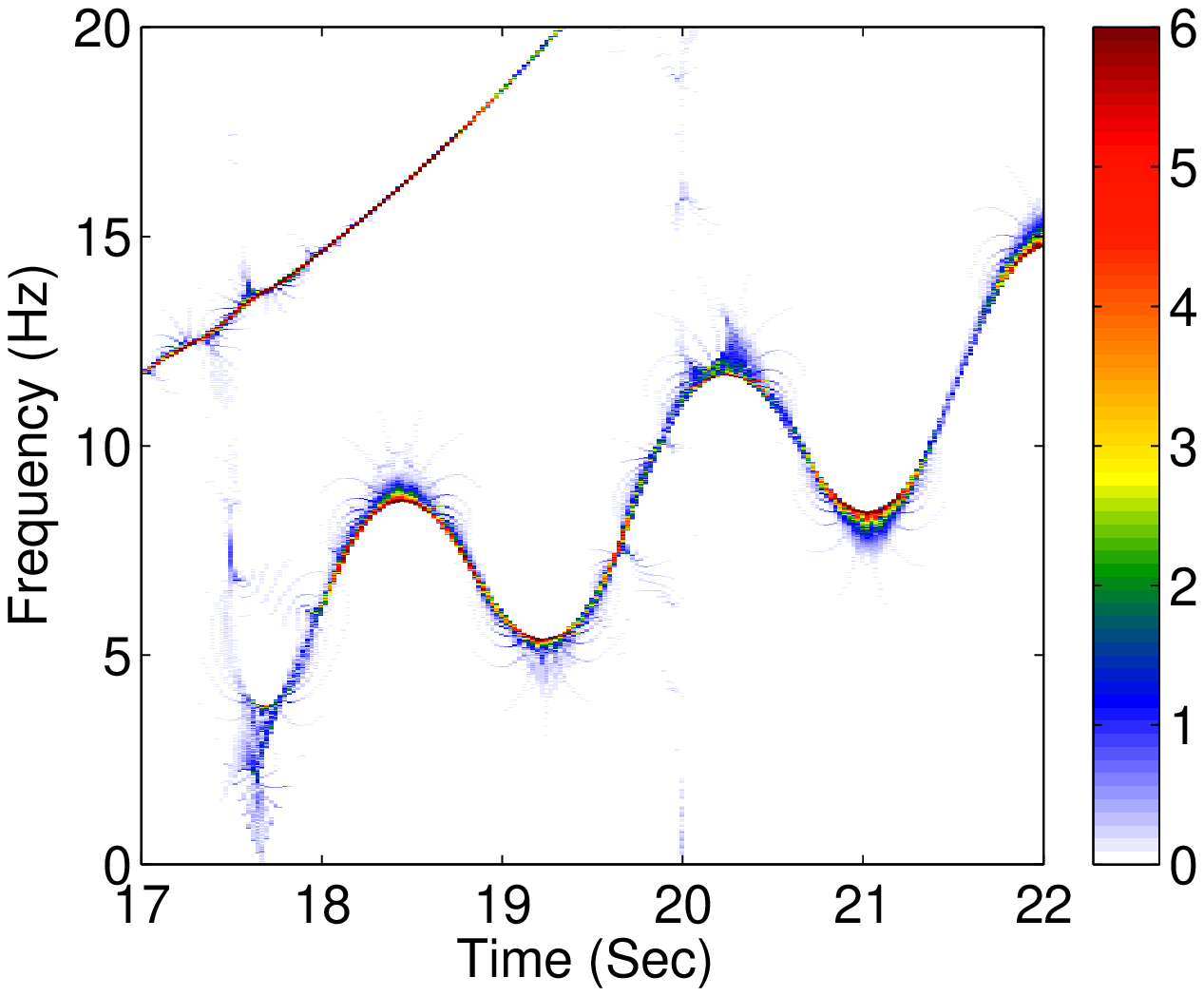}
\caption{oSST, small window width} \label{Symmetry:c}
\end{subfigure}
\begin{subfigure}{0.42\textwidth}
		\includegraphics[width=\linewidth]{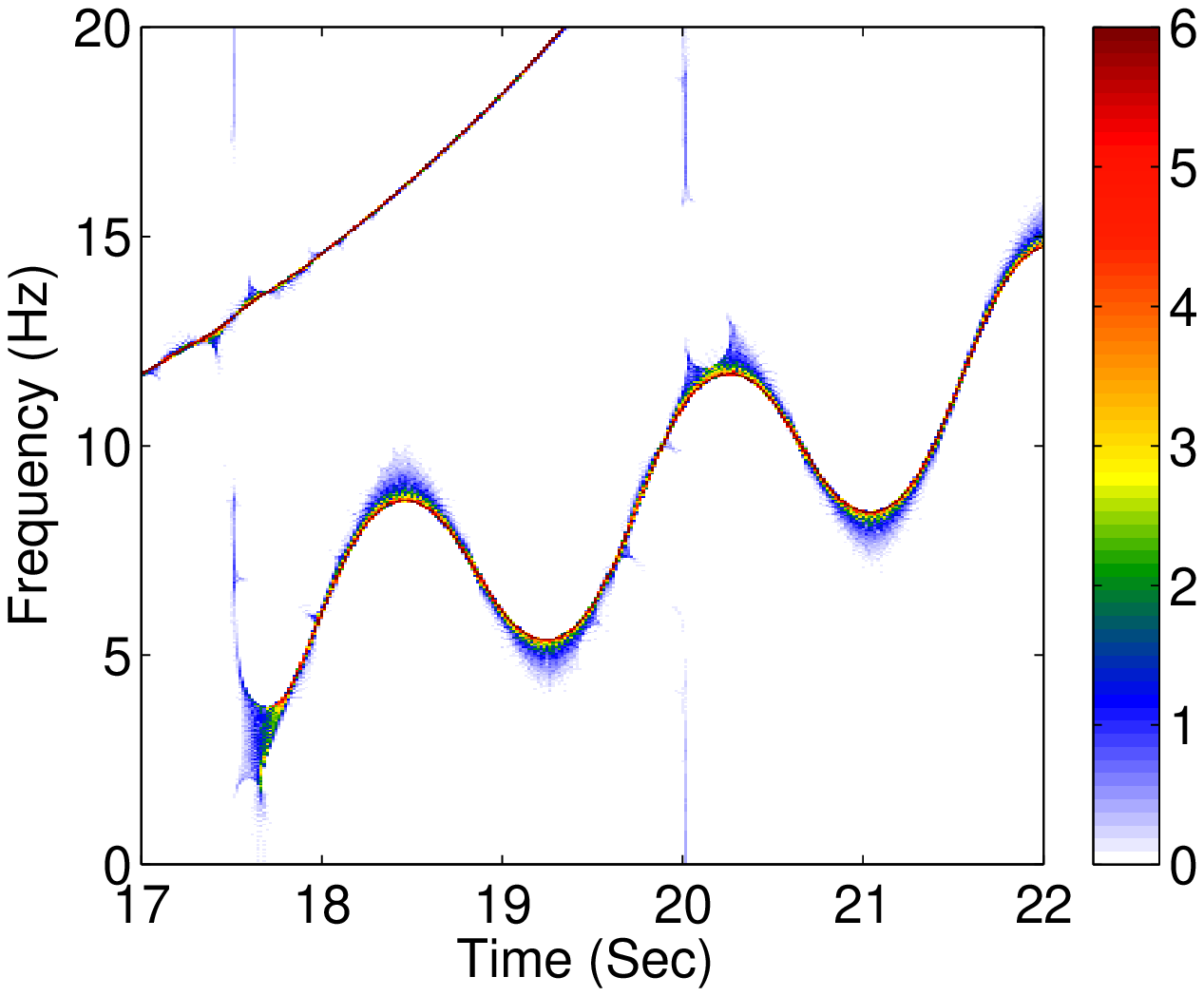}
\caption{RM, small window width} \label{Symmetry:d}
\end{subfigure}

\begin{subfigure}{0.42\textwidth}
		\includegraphics[width=\linewidth]{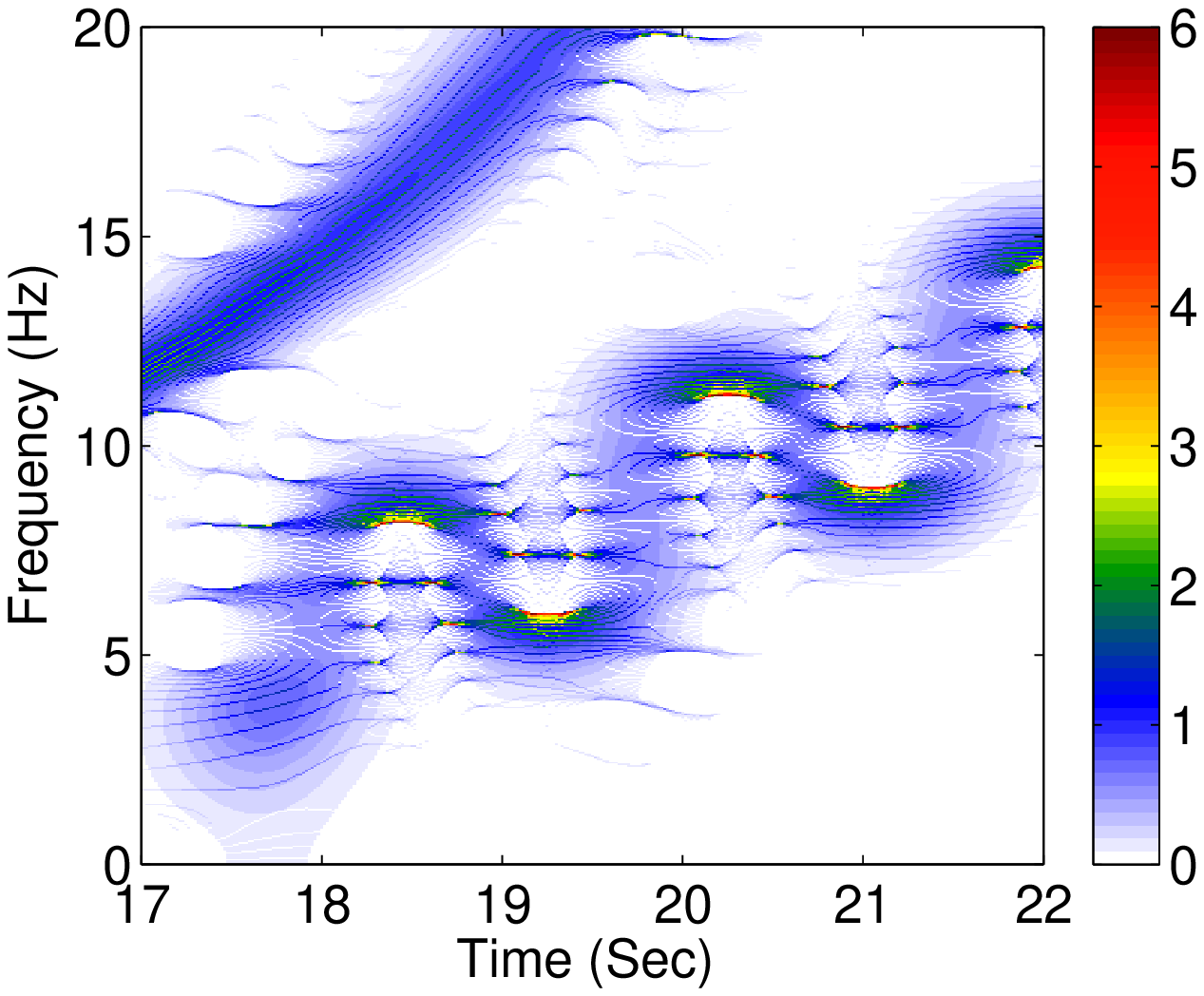}
\caption{SST, large window width} \label{Symmetry:e}
\end{subfigure}
\begin{subfigure}{0.42\textwidth}
		\includegraphics[width=\linewidth]{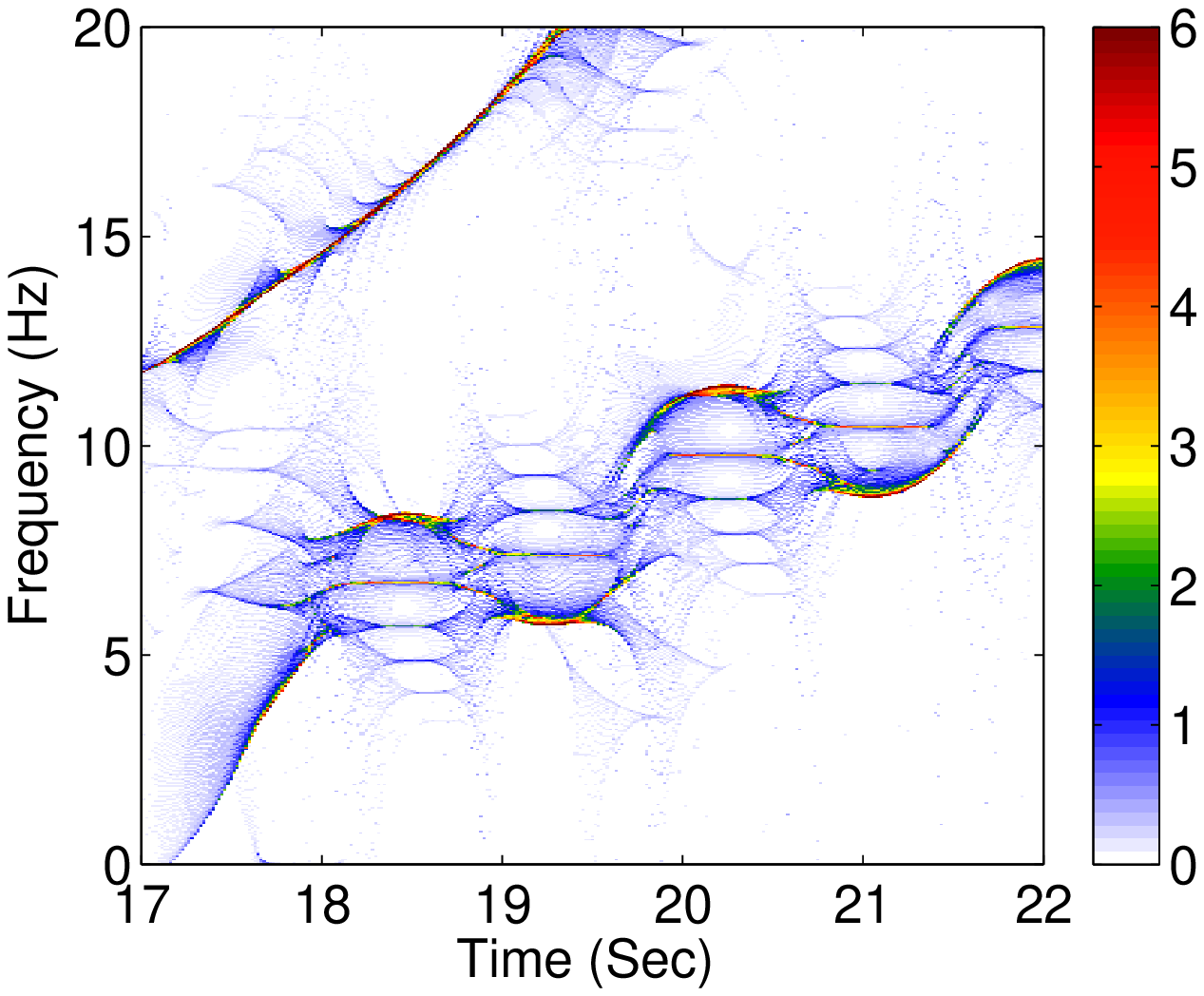}
\caption{vSST, large window width} \label{Symmetry:f}
\end{subfigure}
\begin{subfigure}{0.42\textwidth}
		\includegraphics[width=\linewidth]{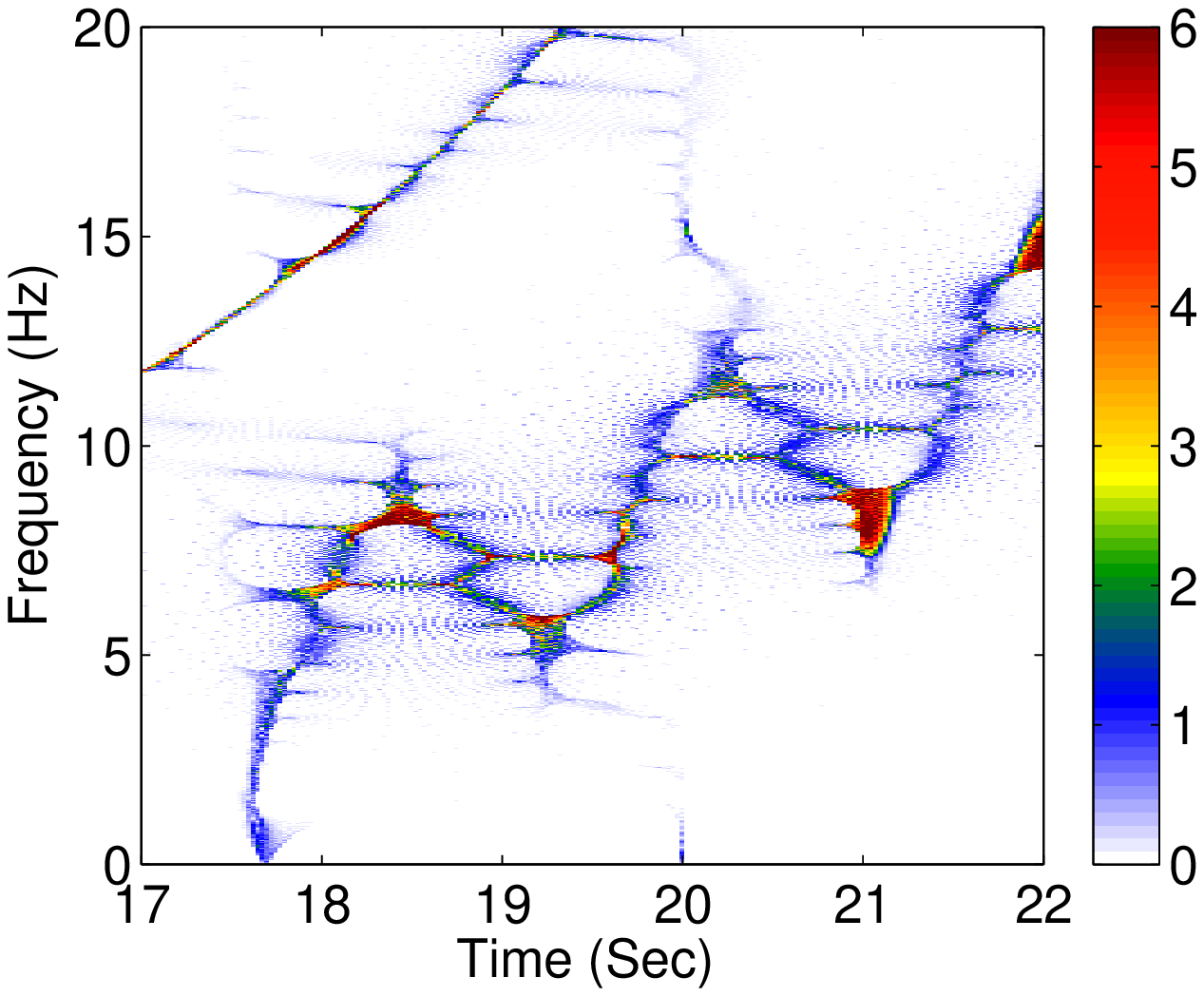}
\caption{oSST, large window width} \label{Symmetry:g}
\end{subfigure}
\begin{subfigure}{0.42\textwidth}
		\includegraphics[width=\linewidth]{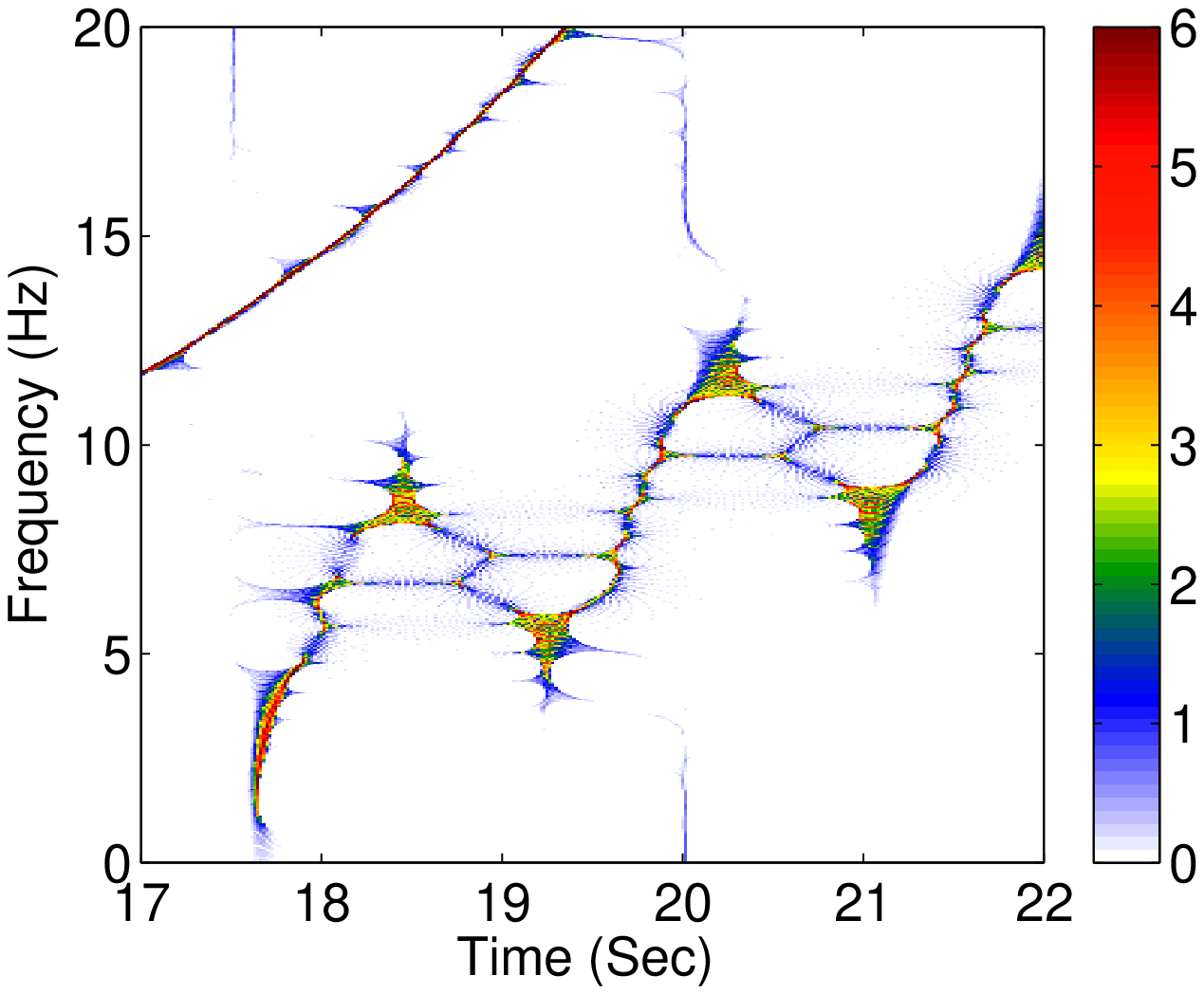}
\caption{RM, large window width} \label{Symmetry:h}
\end{subfigure}

 \caption{\label{Symmetry}A small window width is needed to capture the variation in the oscillatory IF component.
The window width in the upper panel is $\frac{111}{720}$ s (small), and that in the lower panel is $\frac{251}{720}$ s (large).
The TFRs of the SST, the vSST, the oSST, and the RM are shown in (a)(e), (b)(f), (c)(g), and (d)(h), respectively.
The TFR values are normalized by the z-score.
It is clear that while the TFRs of the 2nd-order SST and the RM are sharpened, with the large window width these TFRs are ``confused'' by the fast-varying IFs. A shorter window width in this case can help increase the TFR quality. We could see that the SST could not well handle the fast-varying IF.}
\end{figure}

In this subsection, we accentuate that while the 2nd-order SST and the RM could provide a sharper TFR compared with the SST, the impact of the window width is not negligible.
We demonstrate the TFR of the synthetic signal (\ref{Synthetic}) analyzed by the 2nd-order SST and the RM in Fig.~\ref{Coupling} and Fig.~\ref{Symmetry}. In both figures no noise is involved.
While the 2nd-order SST and the RM can mitigate the limitation of the SST caused by the fast-varying IF components, without a proper choice of the window width, the 2nd-order SST and the RM could fail.

Figure~\ref{Coupling} shows that a large window width is required to separate the two components with closing IFs.
The (a)-(d) in Fig.~\ref{Coupling} are TFRs using a small window width $\frac{111}{720}$ s, which is the GOWW of the vSST.
The coupling artifact between the two components caused by the small width for the all transforms is evident.
As mentioned in previous sections, the coupling artifact can be greatly diminished by increasing the window width.
In (d)-(h) in Fig.~\ref{Coupling}, we choose the window width as $\frac{345}{720}$ s, which is the largest TVOWW for the SST.
For all TFRs, the two IF components are clearly separated, particularly in the RM result Fig.~\ref{Coupling}(h).

Figure ~\ref{Symmetry} shows that a small window width is required to capture the variation in an oscillating IF component.
Note that the small window width is $\frac{111}{720}$ s and the large window width is $\frac{251}{720}$ s.
A small window width provides a fine temporal resolution, which allows us to extract the dynamical information of an IF component ((a)-(d) in Figure ~\ref{Symmetry}), while a large window width causes ambiguity in temporal direction ((e)-(h) in Figure ~\ref{Symmetry}). 

In summary, a proper window width is a prerequisite to obtain the accurate IF information in the TFR in spite of the fact that the conventional reassignment rule in the RM and high-order reassignment rules in the 2nd-order SST can cope with the fast-varying IF components efficiently.

\subsubsection{Reconstruction Error Analysis}

Finally, to quantify the improvement of the TFR by taking the TVOWW into account, we evaluate the normalized root-mean-square deviation (RMSD) by comparing the reconstructed signal components, the IF, and the AM with corresponding true answers.

The normalized RMSD for the evaluation component $\hat{f}_i$, where $i=1$, $2$, and $3$, is given as
\begin{align}
{\rm{normalized \enspace RMSD}}(\hat{f}_i)=\frac{{\|\hat{f}_i^2(t) - f_i^2(t)\|_{L^2}}}{|f_{i,{\rm max}}-f_{i,{\rm min}}|},   \label{nrmsd}
\end{align}
where $f_{i,\rm max}$ and $f_{i,\rm min}$ are the maximum and minimum values of $f_i$, respectively. 
Here $\hat{f}_i$ can represent the reconstructed signal component $\hat{x}_i$, the reconstructed IF and the reconstructed AM from the TFR.

%
%
In addition to the noiseless condition, we compute the normalized RMSD for SNR of $15$ and $10$ dB for $25$ trials, and report the mean and the standard deviation of the normalized RMSD.

\begin{figure*}
\centering

\begin{subfigure}{0.31\textwidth}
		\includegraphics[width=\linewidth]{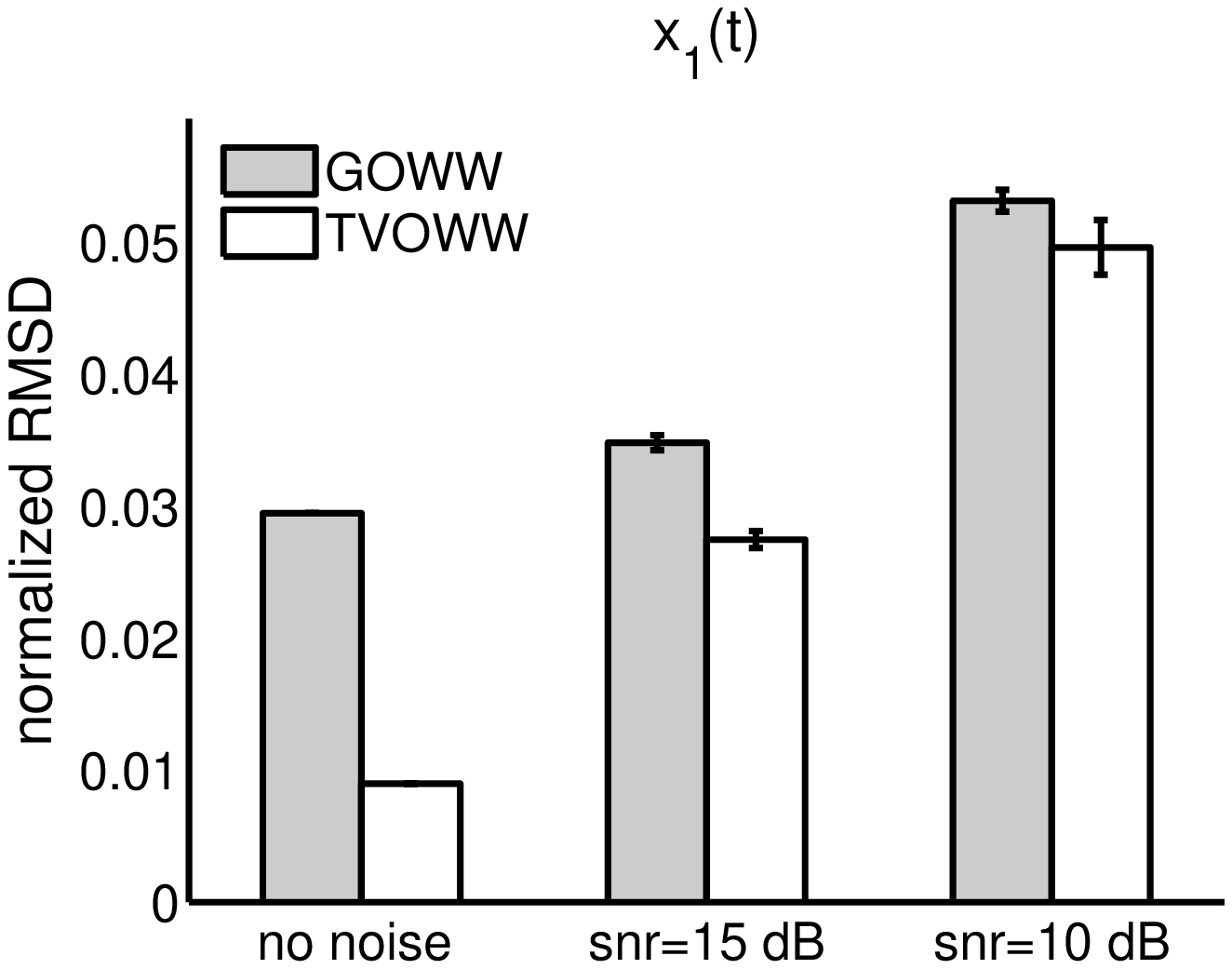}
\caption{} \label{ReconErr:a}
\end{subfigure}
\begin{subfigure}{0.31\textwidth}
		\includegraphics[width=\linewidth]{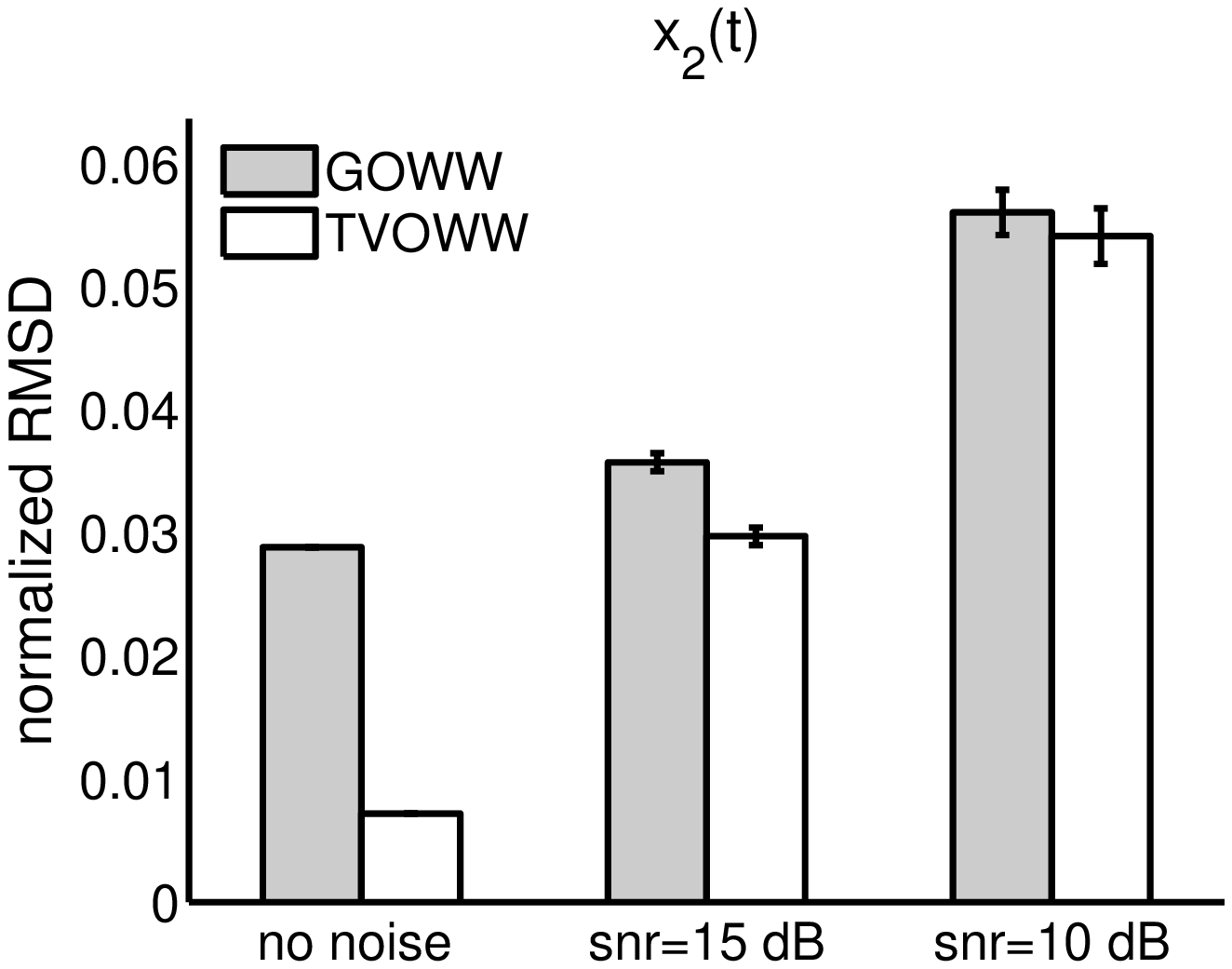}
\caption{} \label{ReconErr:b}
\end{subfigure}
\begin{subfigure}{0.31\textwidth}
		\includegraphics[width=\linewidth]{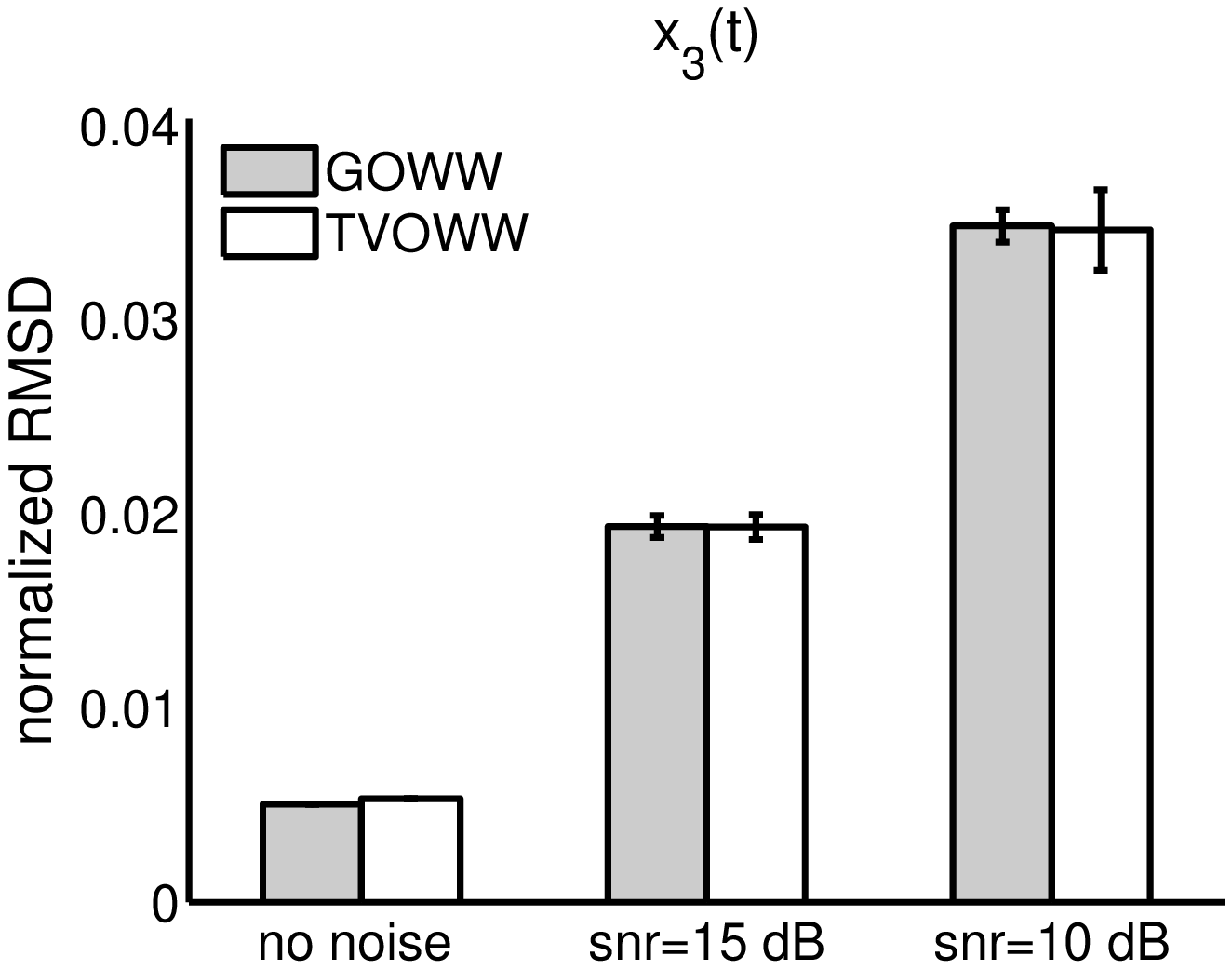}
\caption{} \label{ReconErr:c}
\end{subfigure}
 \caption{The normalized RMSD estimated by comparing the true answer $x_i(t)$ with the reconstructed signal ${\hat x}_i(t)$ from TFRs with the GOWW or the TVOWW for $i=1,2,$ and $3$. Noisy cases are considered for the SNR of $15$ and $10$ dB. 
 }
  \label{ReconErr}
\end{figure*}

\begin{figure*}
\centering

\begin{subfigure}{0.31\textwidth}
		\includegraphics[width=\linewidth]{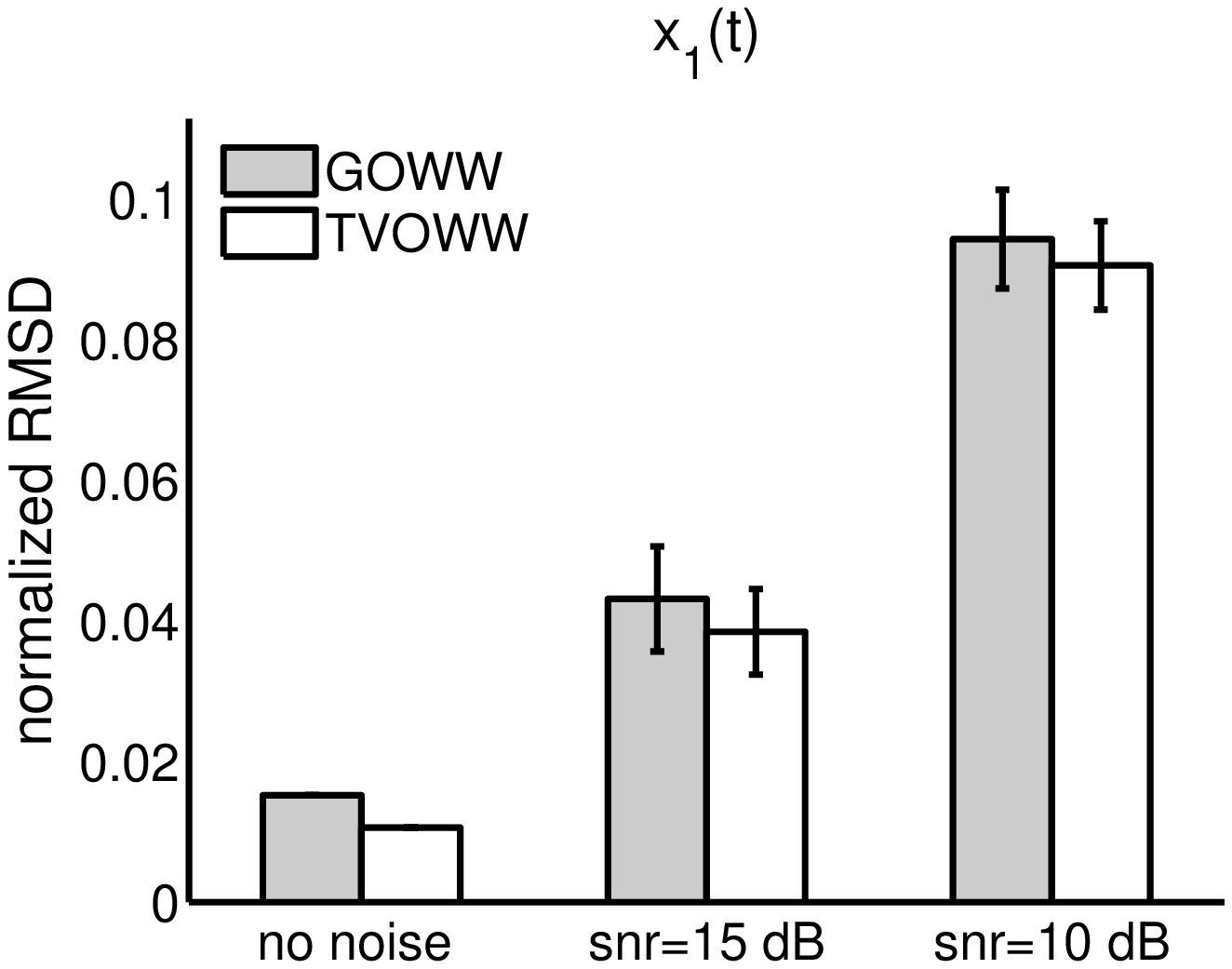}
\caption{} \label{ReconErr:a}
\end{subfigure}
\begin{subfigure}{0.31\textwidth}
		\includegraphics[width=\linewidth]{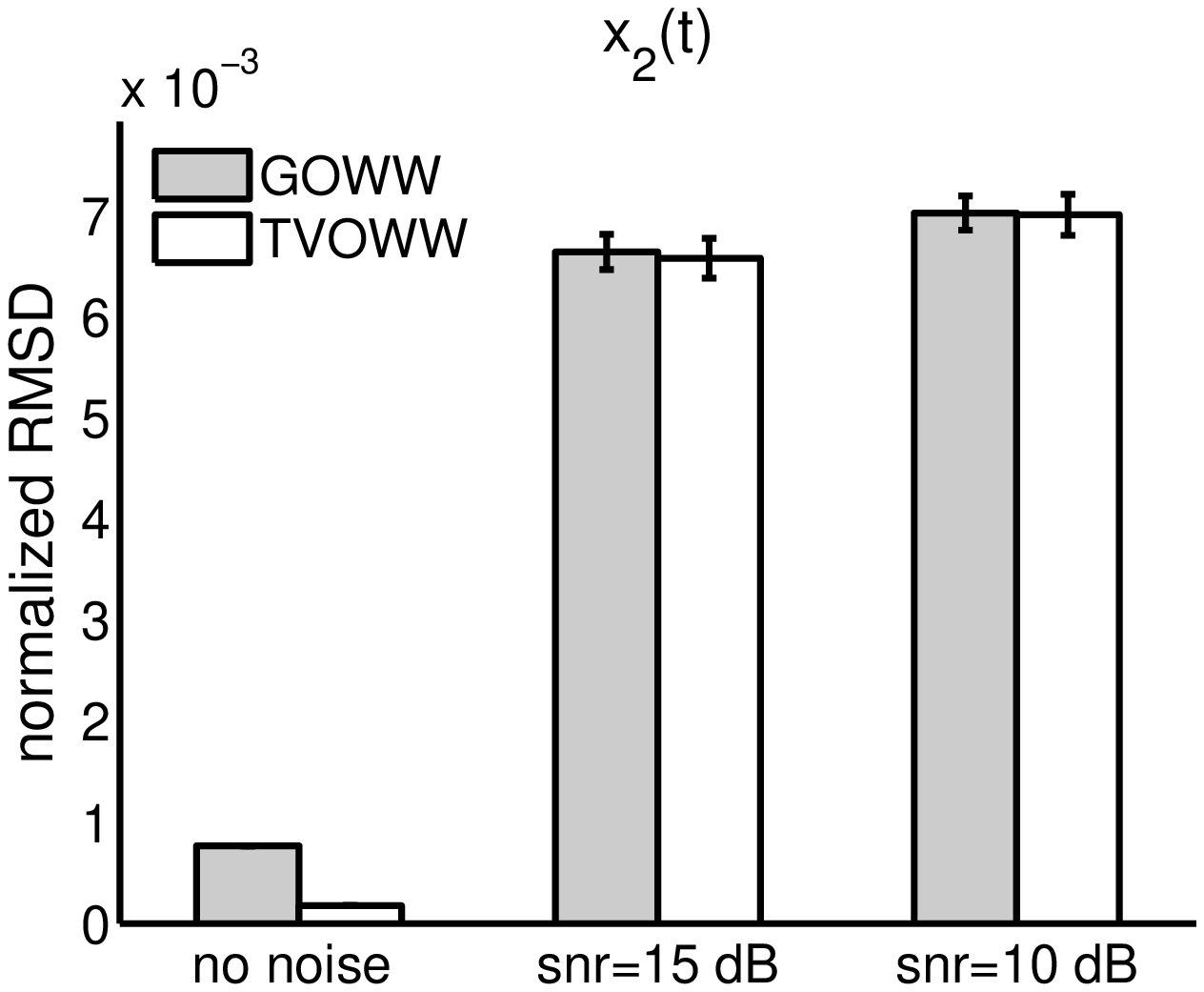}
\caption{} \label{ReconErr:b}
\end{subfigure}
\begin{subfigure}{0.31\textwidth}
		\includegraphics[width=\linewidth]{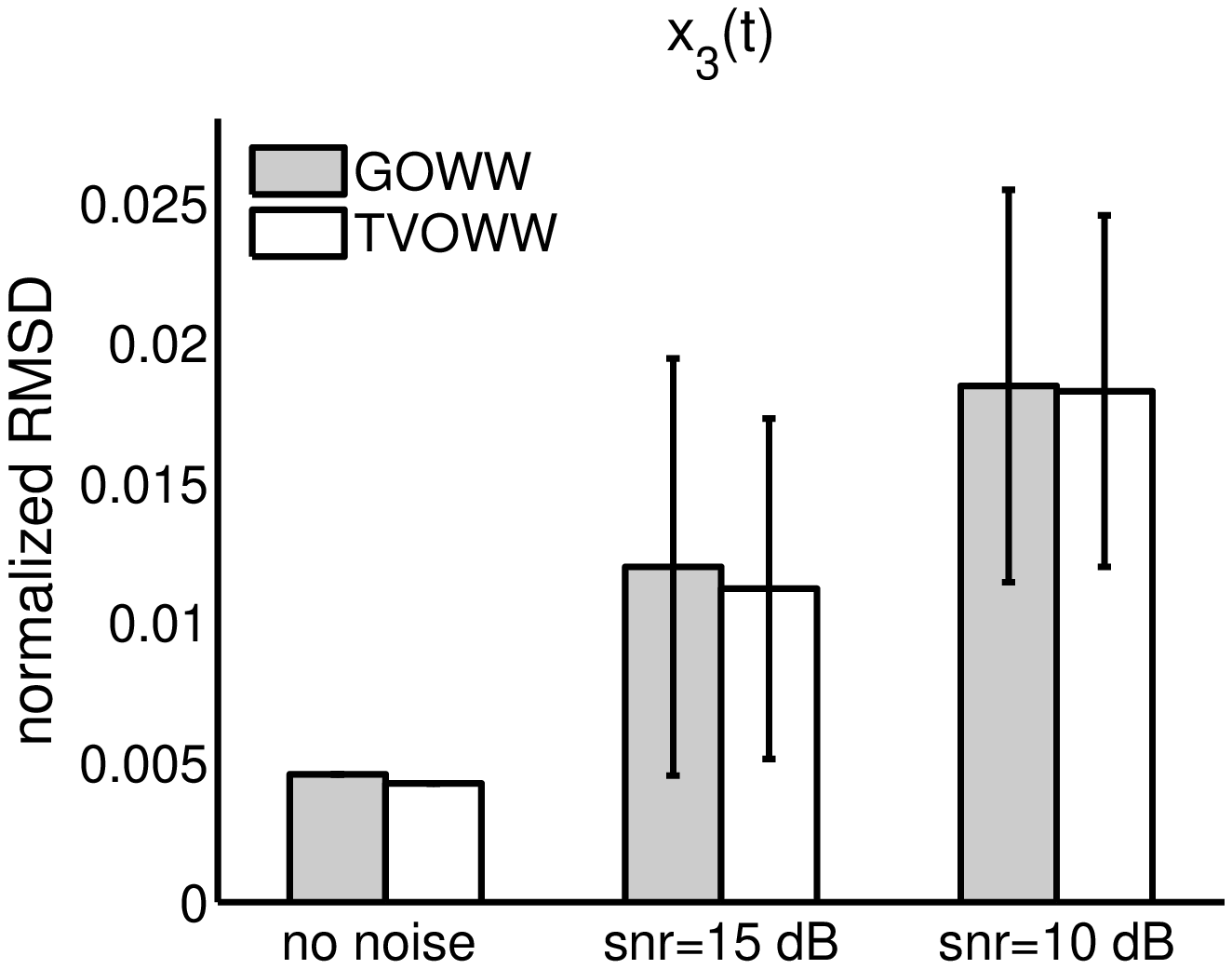}
\caption{} \label{ReconErr:c}
\end{subfigure}
 \caption{The normalized RMSD estimated by comparing the true instantaneous frequency with the reconstructed instantaneous frequency from TFRs with the GOWW or the TVOWW for each IMT. Noisy cases are considered for the SNR of $15$ and $10$ dB.
 }
  \label{ReconErr2}
\end{figure*}

\begin{figure*}
\centering

\begin{subfigure}{0.31\textwidth}
		\includegraphics[width=\linewidth]{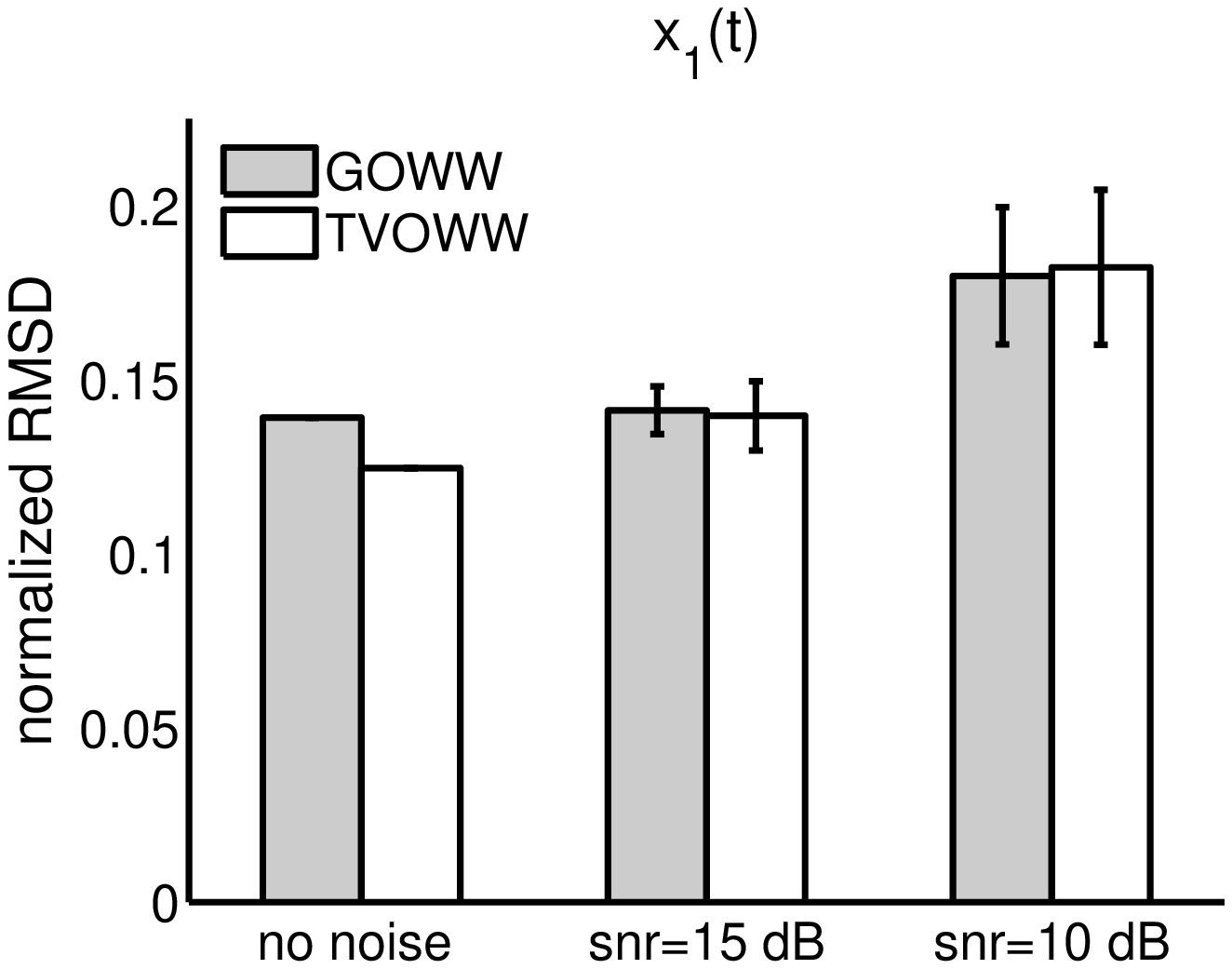}
\caption{} \label{ReconErr:a}
\end{subfigure}
\begin{subfigure}{0.31\textwidth}
		\includegraphics[width=\linewidth]{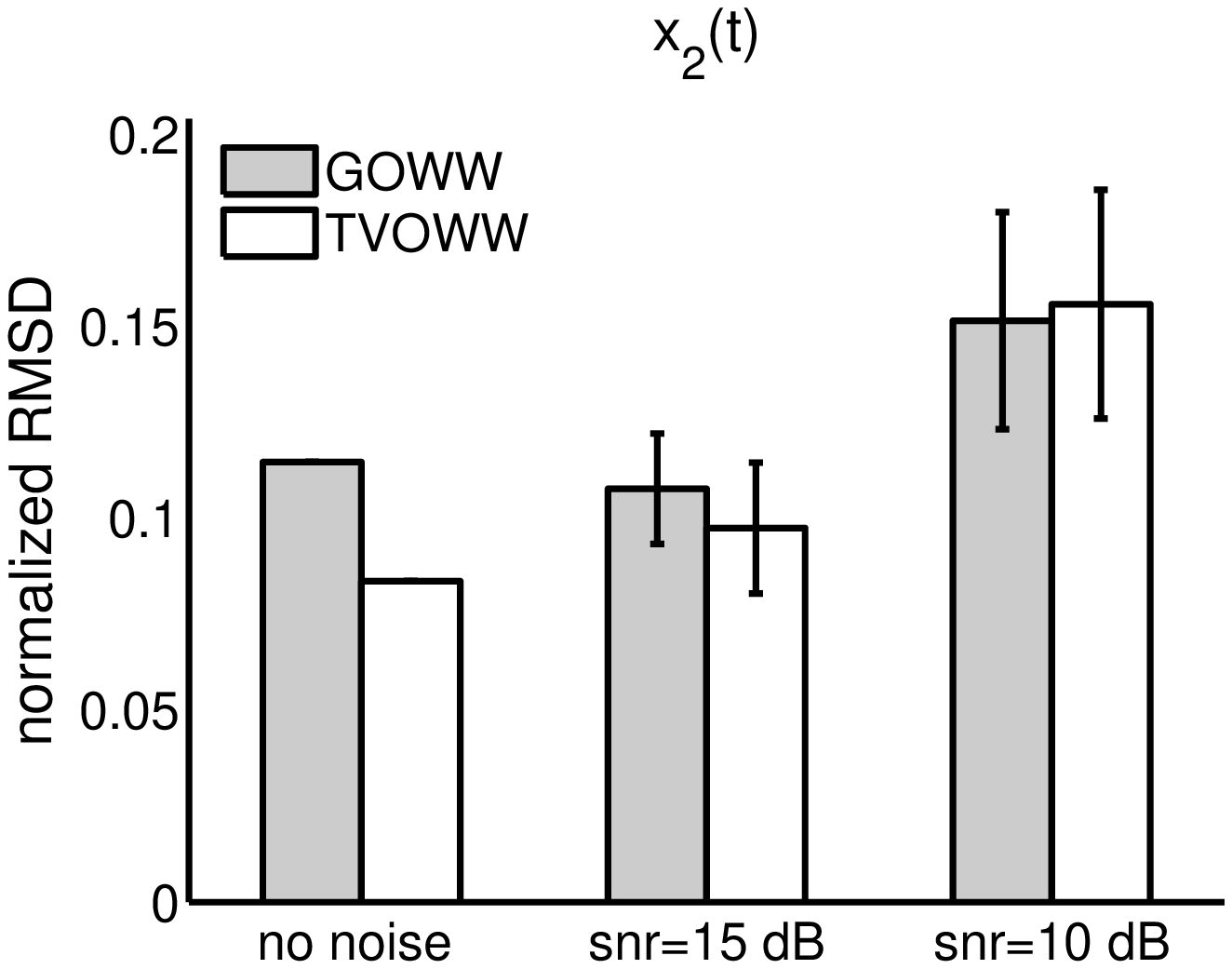}
\caption{} \label{ReconErr:b}
\end{subfigure}
\begin{subfigure}{0.31\textwidth}
		\includegraphics[width=\linewidth]{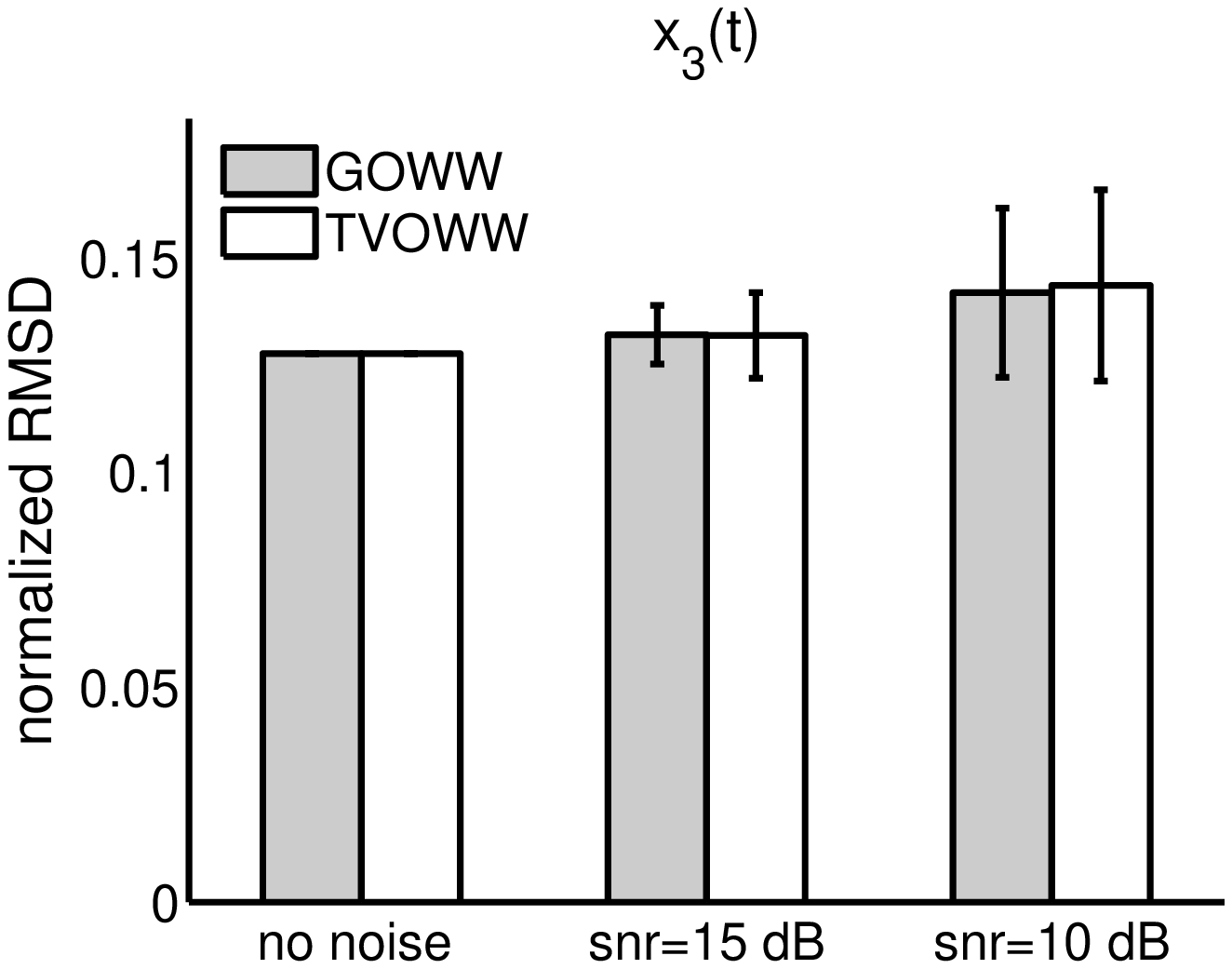}
\caption{} \label{ReconErr:c}
\end{subfigure}
 \caption{{The normalized RMSD estimated by comparing the true amplitude modulation with the reconstructed amplitude modulation from TFRs with the GOWW or the TVOWW for each IMT. Noisy cases are considered for the SNR of $15$ and $10$ dB.} 
 }
  \label{ReconErr3}
\end{figure*}

The results for the reconstruction performance, the reconstructed IF, and the reconstructed AM for each component are presented in Fig.~\ref{ReconErr}, ~\ref{ReconErr2}, and ~\ref{ReconErr3}, respectively. 
%
The IF components are estimated by evaluating the center of mass of the TFR.
The AM components are extracted from the envelope of the reconstructed signal components.
Note that the error varies for different methods to compute the AM components.

The results confirm the benefit of the TVOWW selection scheme, particularly for the components $x_1(t)$ and $x_2(t)$.
For $x_3(t)$, the errors of the GOWW and the TVOWW selection schemes are similar, since this signal component is less coupled with the others.

The results for the reconstructed IF and the reconstructed AM for each component are presented in Figure ~\ref{ReconErr2} and ~\ref{ReconErr3}, respectively.
The IF components are estimated by evaluating the center of mass of the TFR.
The AM components are extracted from the envelope of the reconstructed signal component.
Note that the error varies for different methods to compute the AM components.

Although not shown in the paper due to the page limit, we mention that the TVOWW selection technique can be applied to the 2nd-SST and other variations of the SST to improve the reconstruction quality.

\subsubsection{Toward an Optimally Concentrated TFR -- AOWW}

\begin{figure}
\centering

\begin{subfigure}{0.45\textwidth}
		\includegraphics[width=\linewidth]{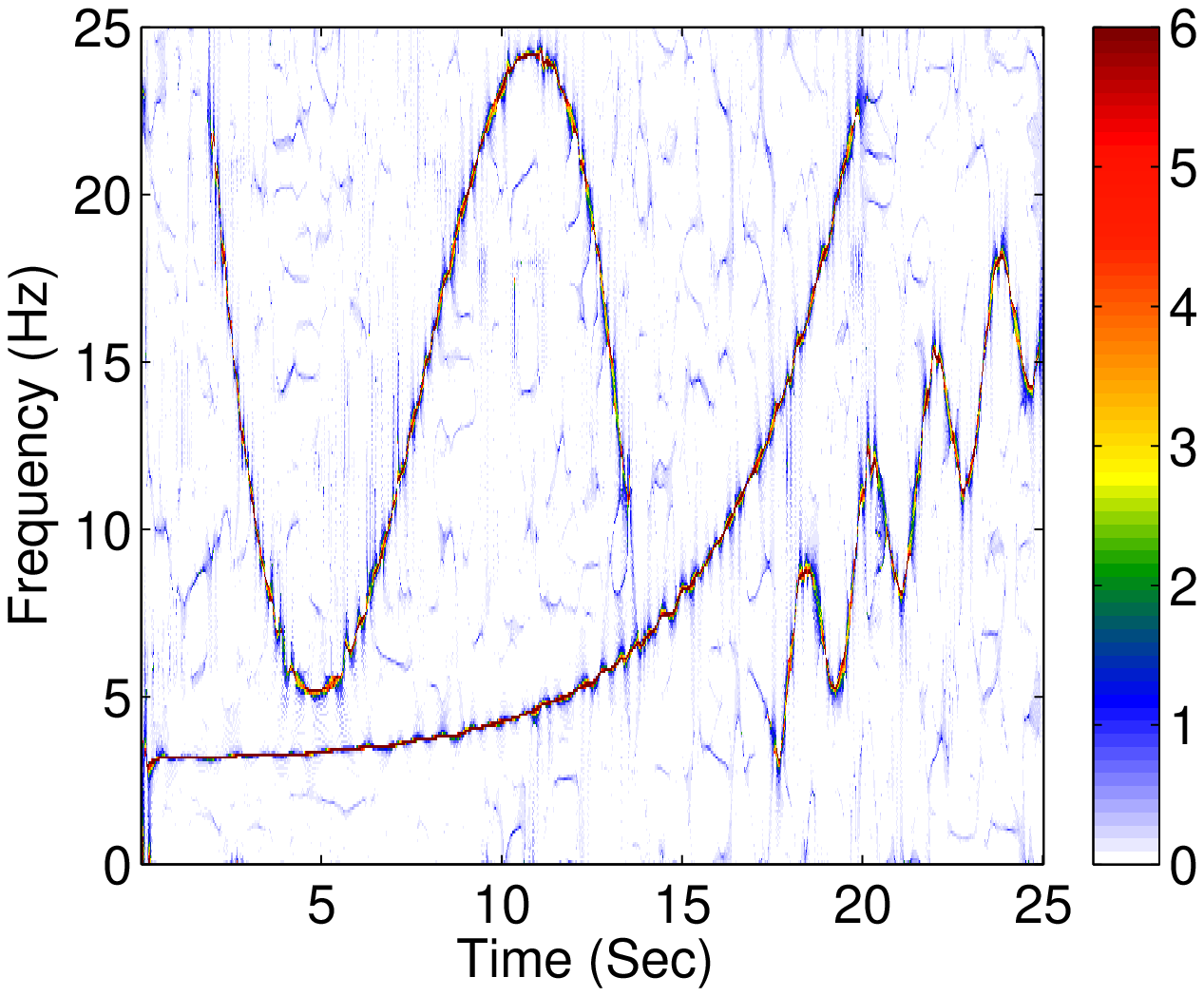}
\caption{SST} \label{AOWW:a}
\end{subfigure}
\begin{subfigure}{0.45\textwidth}
		\includegraphics[width=\linewidth]{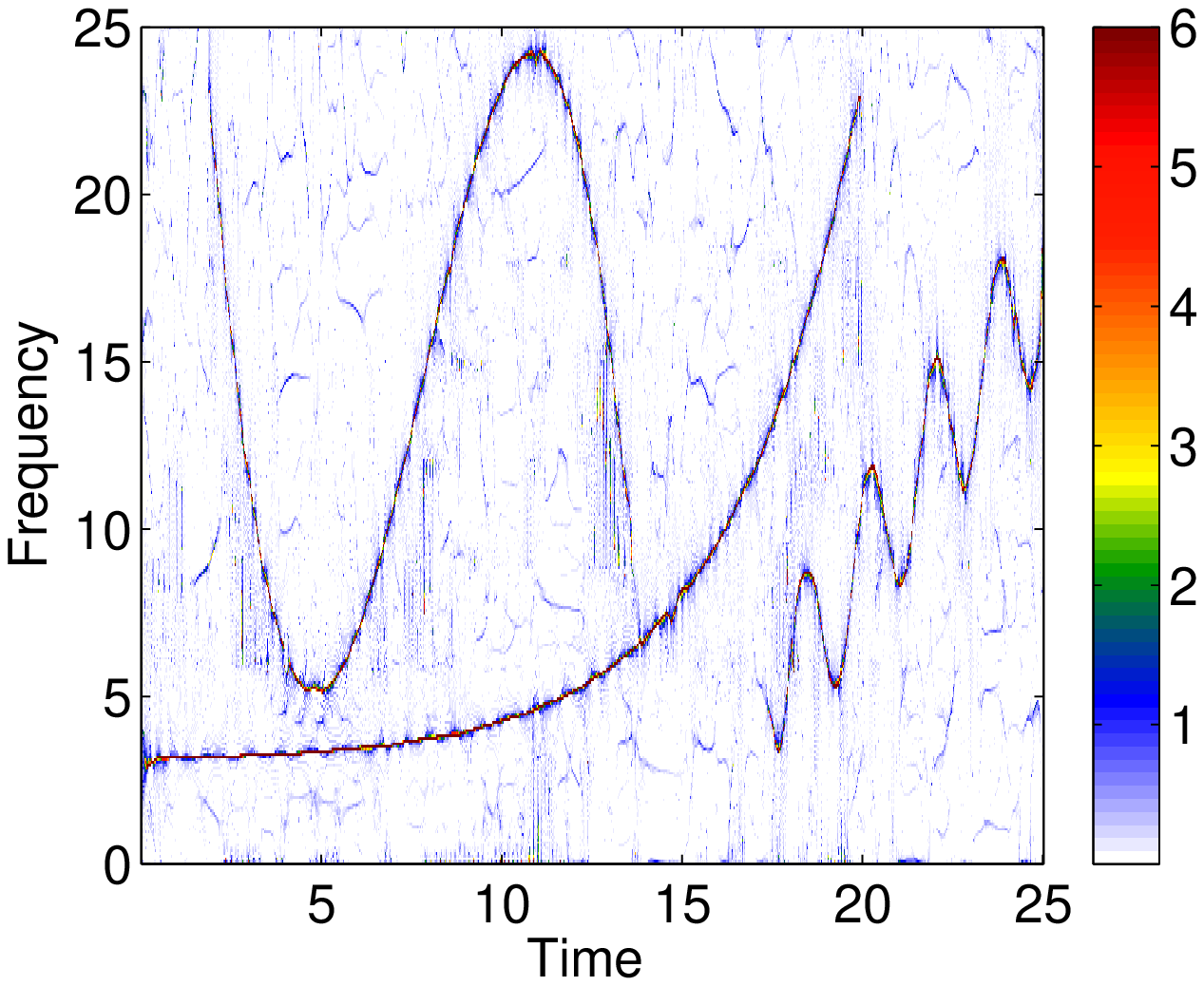}
\caption{vSST} \label{AOWW:b}
\end{subfigure}

 \caption{ The TFRs of the SST and the vSST using the AOWW. }
  \label{AOWW}
\end{figure}

We demonstrate that the AOWW selection scheme can achieve a more concentrated TFR by considering the optimal window width in both time and frequency axes. For the example of the synthetic signal, we set $2b_F=1.6$ Hz and evaluate the local optimal window width every $1.4$ Hz.
The TFR results for the SST and the vSST using the AOWW selection scheme are presented in Fig.~\ref{AOWW}.
By comparing Fig.\ref{AOWW}(a) with Fig.\ref{TVOW}(b), we found that the TFR is sharpened using the AOWW selection scheme, at the expense of significantly increased computation and the lost of the inverse routine to reconstruct each IMT component. Moreover, the TFR of the vSST with the AOWW (Fig.\ref{AOWW}(b)) and that with the TVOWW is similar.

\subsubsection{The Influence of Parameters in the GOWW, the TVOWW and the AOWW Selection Schemes}

We mention that the optimal $\alpha$ value chosen for the Renyi entropy might depend on the application. For a specific application, we could further optimize $\alpha$, and it might depend on parameters such as the sampling rate, frequency-axis and time-axis discretization, and the parameter $b$ and $b_F$.

In this subsection we show that the GOWW, the TVOWW and the AOWW selection schemes are not sensitive to these parameters.
Table 1 presents the normalized RMSD of the reconstructed components for three different $\alpha$, which are $2$, $2.4$, and $2.8$. Here $b$ and $b_F$ are fixed. 
Table 2 presents the normalized RMSD of the reconstructed components for three different $b$, which are $0.8$ s, $0.17$ s, and $0.25$ s. Here $\alpha$ and $b_F$ are fixed.
Since no reconstruction routine is available for the AOWW, the evaluation of the dependence of the AOWW on the chosen parameters is based on the deviation of the IF components via ridge extraction.
Table 3 presents the normalized RMSD of the reconstructed IF components for three different $b_F$, which are $0.6$ Hz, $0.8$ Hz, and $1$ Hz. Here $\alpha$ and $b$ are fixed.
Here the local optimal window width is evaluated every $1$ Hz.
These results provide the evidence that the GOWW, the TVOWW, and the AOWW selection schemes are stable to three major parameters $\alpha$, $b$, and $b_F$.

\begin{table}
\centering
\textbf{Table~1} 
Component reconstruction errors for three different values of $\alpha$ in the GOWW and the TVOWW selection schemes. Here $b$ and $b_F$ are fixed.\\[1ex]
\begin{tabular}{lllll}
        & & $\alpha=2$     & $\alpha=2.4$ &       $\alpha=2.8$ \\
\hline
$x_1(t)$ & GOWW  & $0.1177$ & $0.1177$  &   $0.1177$   \\
         & TVOWW & $0.0357$ & $0.0359$  &   $0.0353$   \\
$x_2(t)$ & GOWW  & $0.1152$ & $0.1152$  &   $0.1152$   \\
         & TVOWW & $0.0288$ & $0.0287$  &   $0.0280$   \\
$x_3(t)$ & GOWW  & $0.0202$ & $0.0202$  &   $0.0202$   \\
         & TVOWW & $0.0210$ & $0.0213$  &   $0.0213$   \\
\end{tabular}
\end{table}

\begin{table}\label{list2}
\centering
\textbf{Table~2} 
Component reconstruction errors for three different values of $b$ in the TVOWW selection scheme. Here $\alpha$ and $b_F$ are fixed.\\[1ex]
\begin{tabular}{llll}
         & $b=0.08s$     & $b=0.17s$ &       $b=0.25s$ \\
\hline
$x_1(t)$ & $0.0354$ & $0.0358$  &   $0.0398$   \\
$x_2(t)$ & $0.0288$ & $0.0294$  &   $0.0342$   \\
$x_3(t)$ & $0.0199$ & $0.0200$  &   $0.0200$   \\
\end{tabular}
\end{table}

\begin{table}\label{list3}
\centering
\textbf{Table~3} 
Instantaneous frequency reconstruction error with three different values of $b_F$ in the AOWW selection scheme. Here $\alpha$ and $b$ are fixed.\\[1ex]
\begin{tabular}{llll}
          & $b_F=0.6$ Hz     & $b_F=0.8$ Hz &       $b_F=1.0$ Hz \\
\hline
$x_1(t)$  & $0.0184$ & $0.0170$  &   $0.0150$   \\
$x_2(t)$  & $3.29\times10^{-4}$ & $2.72\times10^{-4}$  &   $2.66\times10^{-4}$   \\
$x_3(t)$  & $0.0056$ & $0.0049$  &   $0.0050$   \\             

\end{tabular}
\end{table}


\subsection{Application to Attosecond Physics}

During the past decade, real-time observation and direct control of electronic motion in atoms, molecules, nanostructures and solids have been achieved due to advent in the synthesis of attosecond pulses \cite{Krausz_Ivanov:2009}.
In general, an isolated attosecond pulse is created by the superposition of a broadband supercontinuum in high-order harmonic generation driven by high-intensity femtosecond laser pulses \cite{Chini_Zhao_Chang:2014}. To date, an isolated attosecond pulse as short as $67$ attoseconds has been reported \cite{Zhao_Zhang_Chini:2012}. To synthesize shorter attosecond pulses, a better understanding of the underlying physical mechanism is needed. The physical mechanism of the synthesis of attosecond pulses can be understood by analyzing the electron dipole moment oscillation induced by an applied laser field via the TF analysis.
In the previous literature \cite{Antoine_pra:1995,Antoine:1996,Chirila_Dreissigacker__Zwan_Lein_pra:2010,Li_Sheu_Laughlin_Chu:2015,Murakami_Korobkin_Horbatsch_pra:2013,Pfeifer_Gallmann_Abel_Nagel_Neumark_Leone_pra:2006,Sheu_Wu_Hsu:2015,Tong_Chu:2000,Tudorovskaya_Lein:2011}, the linear-type transforms based on short window widths have been adopted and the results are consistent with the classical trajectory simulations \cite{Corkum:1993}. However, there is no discussion on how and why small windows are chosen in the field of attosecond physics.

To clarify this issue, we study the electron dipole moment in atomic hydrogen evoked by an optimally shaped laser waveform that can generate an isolated $21$ attosecond pulse \cite{Chou_Li_Ho_Chu:2015}. 
Such a laser profile can greatly extend the high-order harmonics up to $900$ harmonics within a short time interval, suggesting fast-varying IF components.
The time-dependent dipole moment in the acceleration form is computed by solving a three-dimensional time-dependent Schr{\"o}dinger equation in the framework of the time-dependent generalized pseudospectral (TDGPS) method within the electric dipole approximation \cite{Tong_Chu:1997}. The TDGPS method gives accurate orbital energies and has been employed in the strong field physics as well as attosecond science.
The simulation details are referred to \cite{Chou_Li_Ho_Chu:2015}.

\begin{figure}
\centering
	  \includegraphics[width=0.8\textwidth]{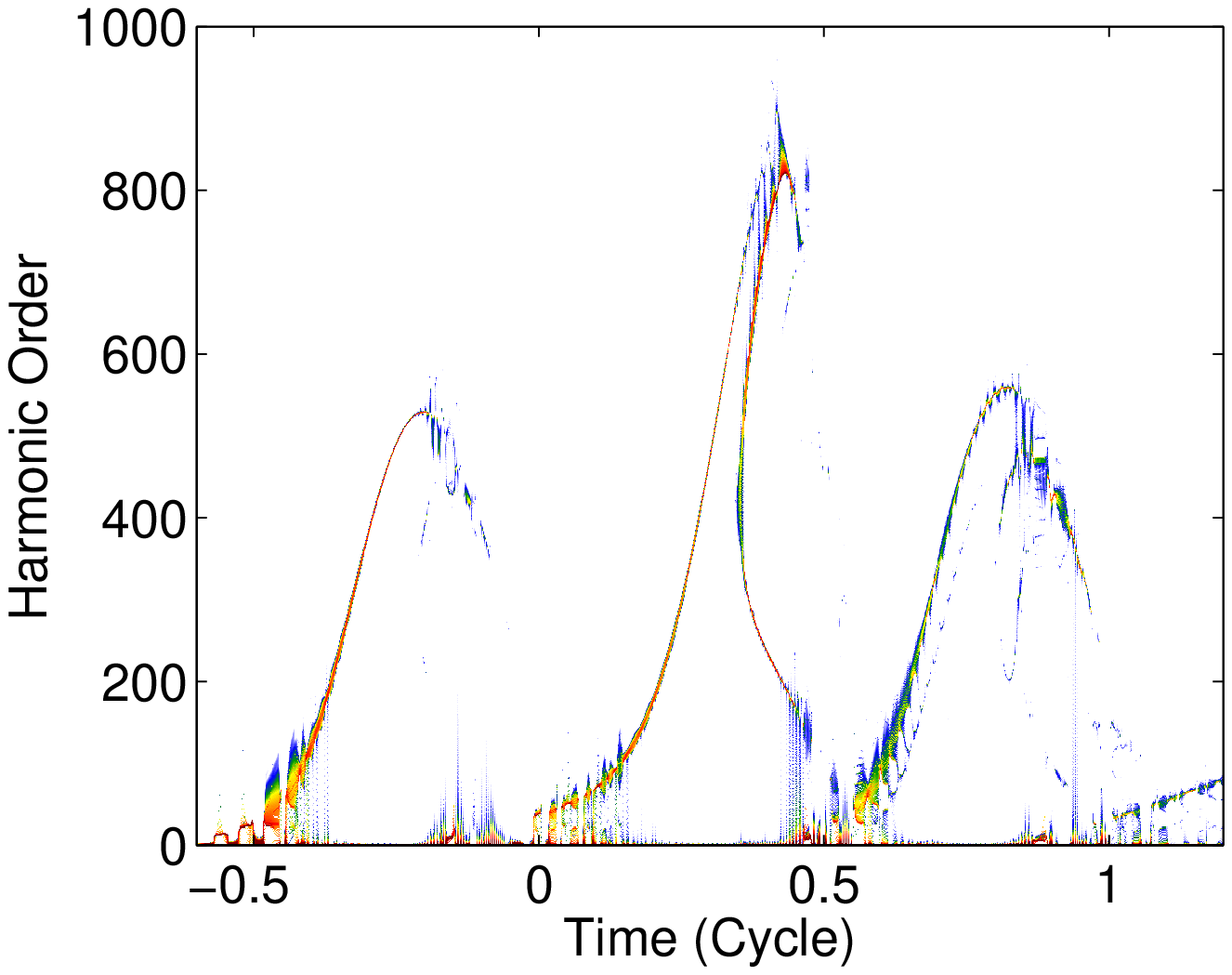}

 \caption{
The TFR of the SST with the TVOWW for the electron dipole moment in an acceleration form. Note that the TFR is in the logarithmic scale. The second emission that reaches up to the $900$th harmonic order in a very short interval can be utilized to synthesize an isolated ultrashort attosecond pulse.
 }
  \label{Chou}
\end{figure}

\begin{figure}
\centering
\begin{subfigure}{0.45\textwidth}
		\includegraphics[width=\linewidth]{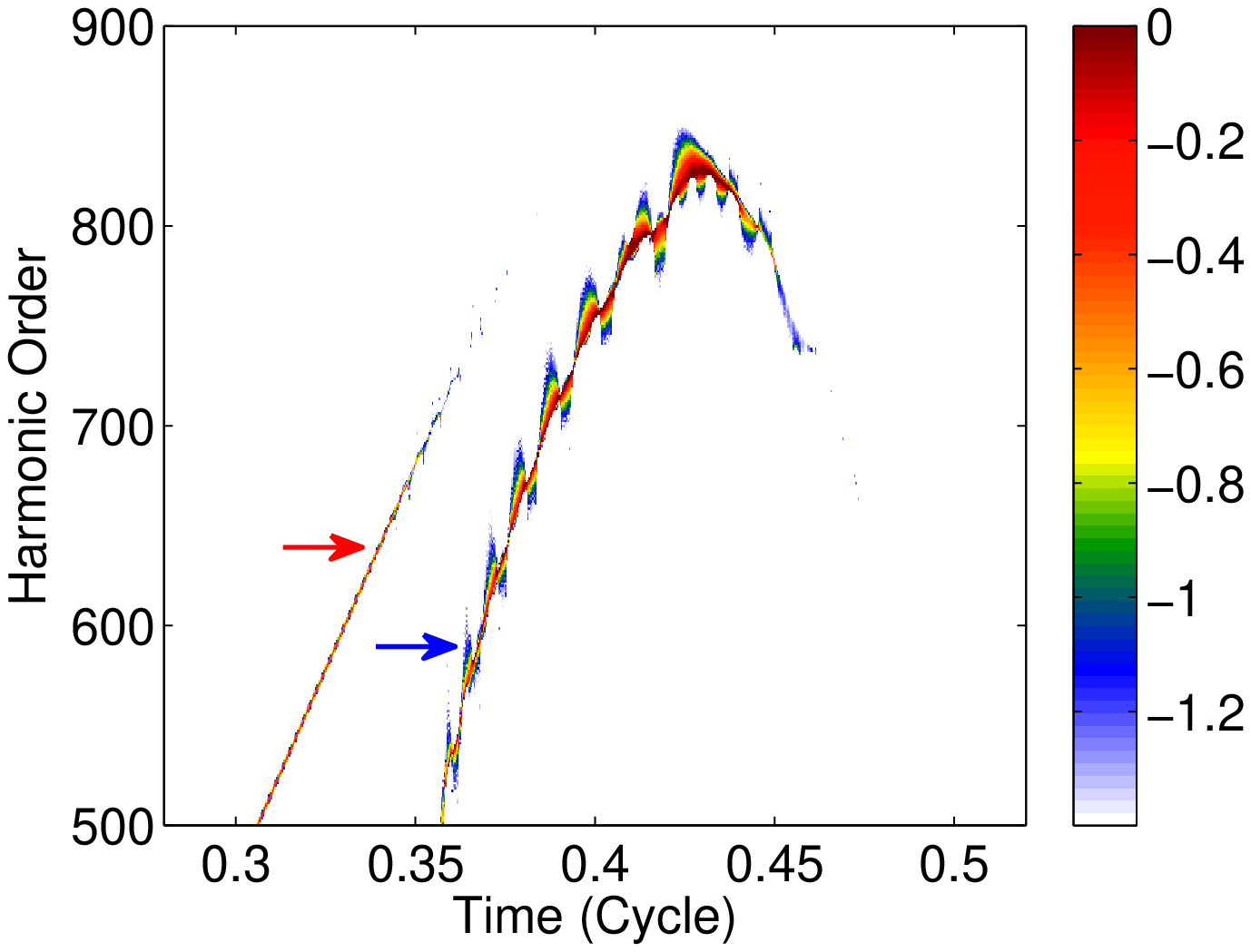}
\caption{GOWW} \label{Chou2:a}
\end{subfigure}
\begin{subfigure}{0.45\textwidth}
		\includegraphics[width=\linewidth]{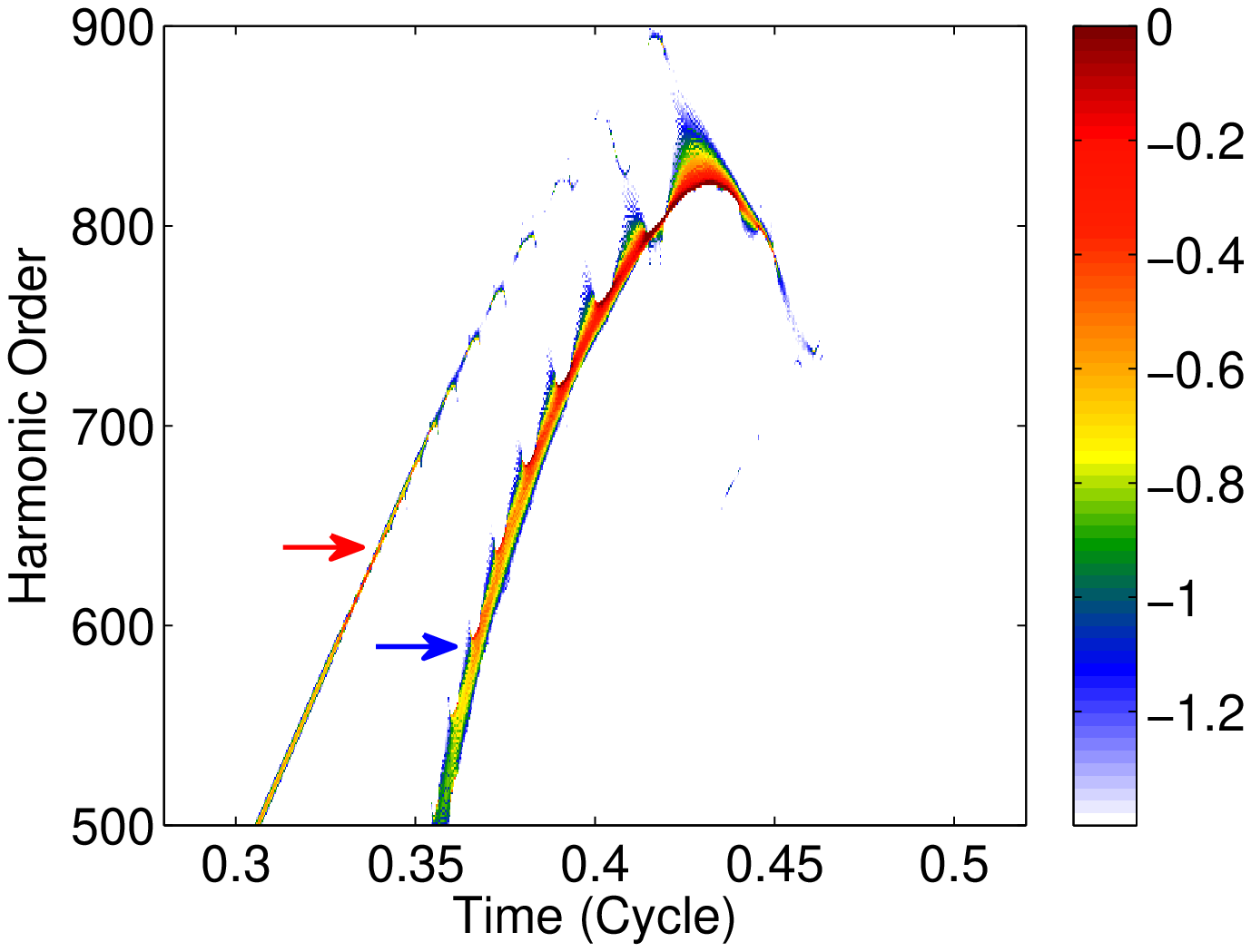}
\caption{TVOWW} \label{Chou2:b}
\end{subfigure}

 \caption{
Enlarged figures from Fig.~\ref{Chou} show delicate differences between the TFR with the GOWW (a) and the TVOWW (b).
The blue arrow indicates the branch corresponding to the long trajectory quantum path and the red arrow indicates the branch corresponding to the short trajectory quantum path.
 }
  \label{Chou2}
\end{figure}

We then compute the R{\'e}nyi entropy for a series of window widths, ranging from $0.25$ atomic units (a.u.) to $8.33$ a.u.
We apply the TVOWW selection scheme with a neighborhood size of $2b=10$ a.u., resulting in Fig.~\ref{Chou}.
Fig.~\ref{Chou} indicates that there are three emissions taking place.
The cutoffs of the first and third emissions are located at around the $500$th order, and the second emission reaches the $900$th order. 
The branches on the TFR nearly coincide with the classical trajectories reported in the previous literature \cite{Chou_Li_Ho_Chu:2015}.
For comparison purposes, enlarged details of the second emission with the GOWW and the TVOWW are displayed in Fig.~\ref{Chou2}.
It is observed that at around $0.43$ laser cycles ($1$ laser cycle $=275.77$ a.u.), the branch indicated by the blue arrow in Fig.~\ref{Chou2}(b) corresponding to the long trajectory quantum path has the strongest intensity and consists of the most harmonics.
While the branch indicated by the red arrow dies out after $0.35$ laser cycles in the TFR of the SST with the GOWW (Fig.~\ref{Chou2}(a)), it is revealed by the result with the TVOWW (Fig.~\ref{Chou2}(b)) that the short trajectory quantum path also has an influence on the high order harmonic emission.
These high order harmonics occur almost simultaneously, which is a prerequisite of a dependable attosecond pulse.

\begin{figure}
\centering
\begin{subfigure}{0.45\textwidth}
		\includegraphics[width=\linewidth]{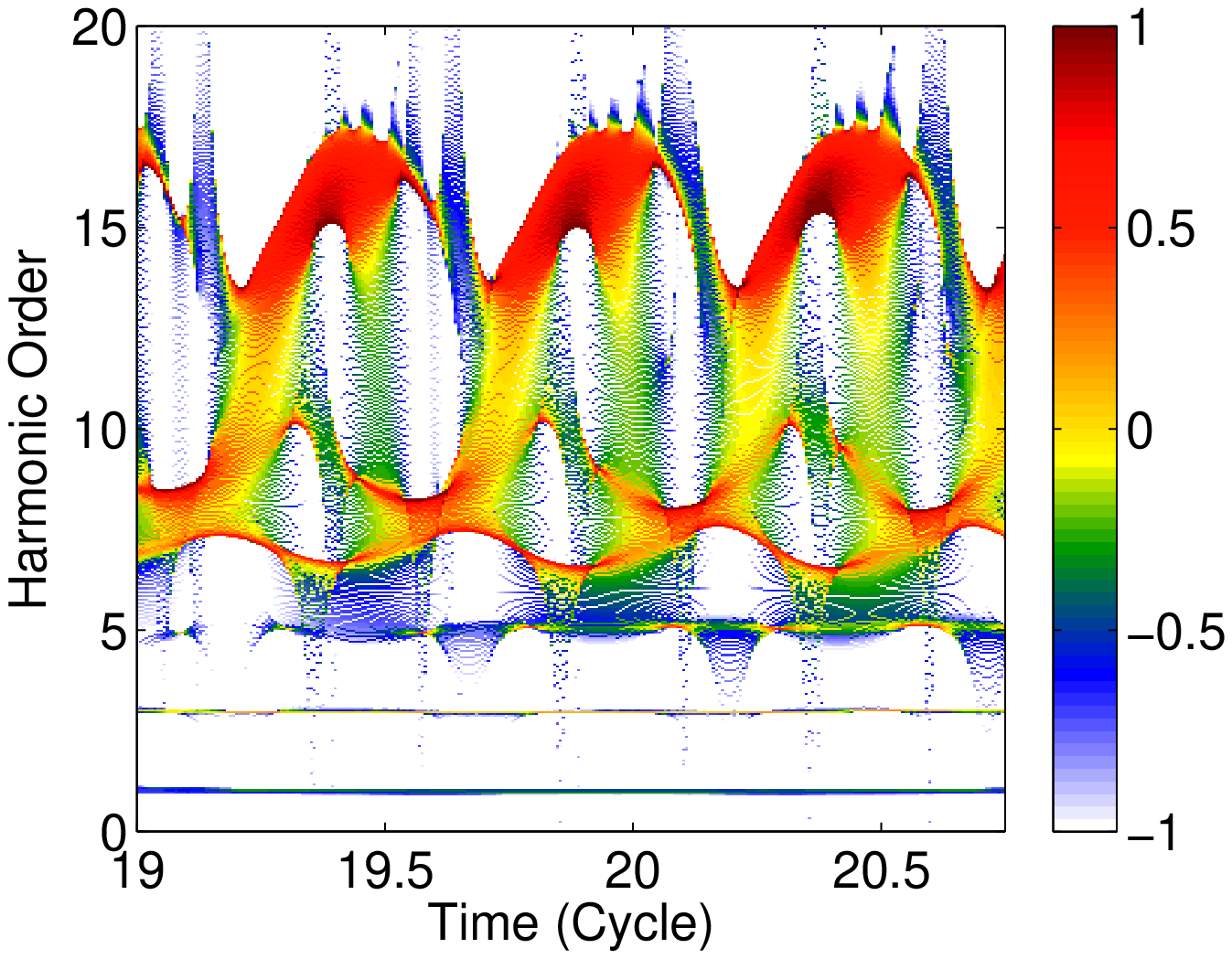}
\caption{Synchrosqueezed Morlet wavelet transform} \label{Risoud:a}
\end{subfigure}
\begin{subfigure}{0.45\textwidth}
		\includegraphics[width=\linewidth]{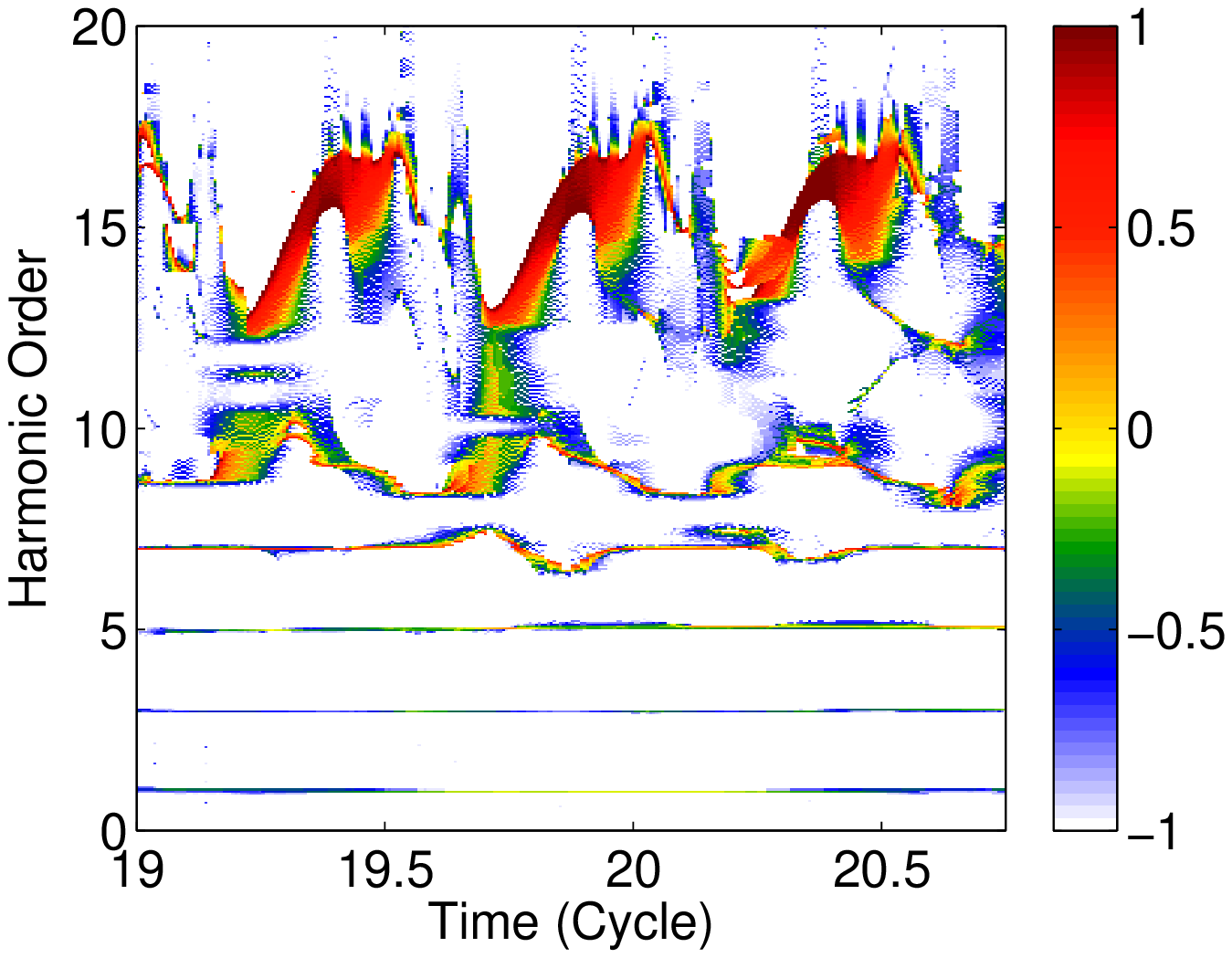}
\caption{AOWW applied to the Synchrosqueezed Morlet wavelet transform} \label{Risoud:b}
\end{subfigure}

 \caption{
(a) The TFR of the synchrosqueezed Morlet wavelet transform of the acceleration dipole moment using a laser field of a wavelength of $800$ nm and an intensity of $5\times10^{13}\;\mbox{W/cm}^2$.
(b) The TFR of the synchrosqueezed Morlet wavelet transform with the AOWW selection scheme applied.
 }
  \label{Risoud}
\end{figure}

In the second example, we demonstrate that the AOWW selection scheme is beneficial to distinguish the near-threshold harmonics in the TF representation of a hydrogen atom in the strong laser field .
Figure \ref{Risoud} shows the TF representations for HHG generated by a monochromatic laser field with a wavelength of $800$ nm and an intensity of $5\times10^{13}\;\mbox{W/cm}^2$.
The laser field profile is described by $\sin^2 (\pi t/(nT))$, where $n=40$ is the pulse length measured in optical cycles ($T=2\pi/\omega_0$), and $\omega_0$ is fundamental angular frequency of the laser wavelength.
(The definition of the laser field profile and the simulation details can be found in \cite{Sheu_Wu_Hsu:2015,Sheu_Aip:2014}.)
The laser parameters correspond to the Keldysh parameter $\gamma_K=1.51$ \cite{Chang:2011,Sheu_Wu_Hsu:2015}, indicating that the main dynamic mechanism is the multiphoton ionization process.
Generally speaking, $\gamma_K\gg1$ and $\gamma_K\ll1$ correspond to the multiphoton ionization regime and the tunneling ionization regime, respectively.
Figure \ref{Risoud}(a) presents the result of the synchrosqueezed Morlet wavelet transform with a scaling parameter $\tau=6$, and in Fig.~\ref{Risoud}(b) the AOWW selection scheme is applied.
Due to the advantage of multiresolution, the synchrosqueezed Morlet wavelet transform \cite{Li_Sheu_Laughlin_Chu:2015,Sheu_Wu_Hsu:2015} can clearly describe the below-threshold harmonics (from the $1$st to the $5$th harmonics), and the chirp-like dynamics in the above-threshold region.
However, in the near-threshold region (The ionization threshold in this case is the $8.78$th harmonic.), the harmonics (i.e., from the $7$th to the $11$th harmonics) are coupled and ambiguous.
After applying the AOWW selection scheme with a neighborhood size of $2b=0.24 T$ and $2b_F=0.46 \omega_0$, 
the near-threshold harmonics in Fig.~\ref{Risoud}(b) are clearly depicted in the TFR.
The second example indicates that the AOWW selection scheme may be applied to other atomic systems such as the Cs atom \cite{Li_Sheu_Laughlin_Chu:2015}.


\subsection{A comparison with other methods}

The proposed TVOWW and the AOWW selection schemes have similarities with some non-reassignment-type TF analysis methods. 
For example, in the sparsification approach \cite{Hou_Shi:2013a}, when the signal satisfies the regularity conditions of the AHM, a dictionary design and a sparsity based optimization lead to the desired time-varying spectral information and signal decomposition. However, it is not clear how to achieve the optimal dictionary design, and the optimization step in the sparsification could be compute-intensive if the dictionary is chosen improperly. To have a parallel comparison with the reassignment-type transforms, note that the dictionary in the reassignment-type transforms, for example, $\mathcal{D}$ in (\ref{Definition:RedundantDictionary}), is infinitely redundant. The ``optimal'' frame is not chosen by any direct optimization procedure. Instead, the reassignment rule provides an approximation of the optimal frames. When combined with the TVOWW or the AOWW selection scheme, we get the optimal frame over an infinitely redundant dictionary. In this sense, when combined with the TVOWW or the AOWW, the reassignment-type transforms could be viewed as a variation of the sparsification approach. 

The Tycoon \cite{Kowalski_Meynard_Wu:2015}, on the other hand, could be viewed as a TF analysis technique based on the convex optimization from the synthesis viewpoint \cite{Balazs_Doerfler_Kowalski_Torresani:2013}. In this approach, we do not design a dictionary or choose a window. Instead, we need to determine some fundamental quantities that a ``good'' TFR should satisfy, and then directly find this good TFR by optimizing a functional capturing the determined fundamental quantities. Since the TFR determined by the SST could approximate the considered functional \cite{Daubechies_Lu_Wu:2011,Kowalski_Meynard_Wu:2015}, the combination of the TVOWW or the AOWW and the SST and its variations could be viewed as a relaxation of the Tycoon.

In the TFJP \cite{Jaillet_Torresani:2007}, we first fix a TF plane tiling. For each block in the TF plane tiling, the optimal window for the GT is then selected based on the R\'enyi entropy. While it leads to a sharper TFR, the ``uncertainty'' still exists. Furthermore, since the TF plane tiling is not uniformly distributed, the signal decomposition ability is limited. 
While the SST and its variations combined with the TVOWW or the AOWW selection schemes could be viewed as a variation of the TFJP in the sense of ``window selection'', we mention that the TFJP and frame-based methods are different in essence. Specifically, TFJP is not specifically designed for sums of frequency modulated signals but for a more general signal, so the application fields of the TFJP are different.

\section{Conclusions}\label{Section:Conclusions}

In this study, we propose two optimal window width selection schemes, namely, the TVOWW and the AOWW selection techniques, to optimize the concentration of the TFR determined by a chosen TF analysis. The R{\'e}nyi entropy is applied to determine the concentration of the TFR.
In addition to showing the performance of the proposed scheme in a synthetic signal, we show potential applications of this method to attosecond physics.
We believe that this work can serve as a cornerstone in ultrafast dynamics in atoms and molecules to uncover new physics.

\section{Acknowledgement}
Hau-tieng Wu acknowledges the support of Sloan Research Fellow FR-2015-65363 and Professor Bruno Torresani and Professor Matthieu Kowalski for the fruitful discussions for different topics related to this work. 
Yae-lin Sheu thanks Prof. S.-I Chu and Y. Chou for their advice on numerical parameters of the generalized pseudospectral method. The authors thank the anonymous reviewers for their constructive suggestions to improve the manuscript.

\bibliographystyle{plain} 
\bibliography{TFanalysis}


\end{document}